\documentclass[a4paper]{aa}
\usepackage{txfonts,graphicx,natbib,aalongtable}
\bibpunct{(}{)}{;}{a}{}{,}
\newcommand{\cz}{\ensuremath{C_Z}}
\newcommand{\pv}{\ensuremath{P_V}}
\newcommand{\nv}{\ensuremath{N_V}}

\newcommand{\bz}{\ensuremath{\langle B_z\rangle}}
\newcommand{\nz}{\ensuremath{\langle N_z\rangle}}
\newcommand{\sbz}{\ensuremath{\sigma_{\langle B_z\rangle}}}
\newcommand{\snz}{\ensuremath{\sigma_{\langle N_z\rangle}}}
\newcommand{\fo}{\ensuremath{f^\parallel}}
\newcommand{\fe}{\ensuremath{f^\perp}}

\begin{document}
\title{The FORS1 catalogue of stellar magnetic field measurements\thanks{The full catalogue and the spectra are available in electronic form
at the CDS via anonymous ftp to cdsarc.u-strasbg.fr (130.79.128.5)
or via http://cdsweb.u-strasbg.fr/cgi-bin/qcat?J/A+A/. This paper includes an abridged
printable version of the catalogue.}}
       \author{
        S.~Bagnulo      \inst{1}
       \and
        L.~Fossati      \inst{2,3}
       \and
        J.D.~Landstreet  \inst{1,4}
       \and
        C.~Izzo          \inst{5}
        }
\institute{
           Armagh Observatory,
           College Hill,
           Armagh BT61 9DG,
           Northern Ireland, U.K.
           \email{sba@arm.ac.uk, jls@arm.ac.uk}
           \and
           Space Research Institute,
           Austrian Academy of Sciences,
           Schmiedlstrasse 6, 
           A-8042 Graz, Austria.\\
           \email{luca.fossati@oeaw.ac.at}
           \and
           Argelander Institut f\"{u}r Astronomie, 
           Auf dem H\"{u}gel 71, 
           Bonn D-53121, Germany.
           \and
           Physics \& Astronomy Department,
           The University of Western Ontario,
           London, Ontario, Canada N6A 3K7. \\
           \email{jlandstr@uwo.ca}
           \and
           Deceased
}
\authorrunning{S.\ Bagnulo et al.}
\titlerunning{The FORS1 catalogue of stellar magnetic field measurements}
\date{Received: 2015-05-11 / Accepted: 2015-08-03}
\abstract
{
The FORS1 instrument on the ESO Very Large Telescope was used to
obtain low-resolution circular polarised spectra of nearly a thousand
different stars, with the aim of measuring their mean longitudinal
magnetic fields. Magnetic fields were measured by different authors,
and using different methods and software tools.
}
{
A catalogue of FORS1 magnetic measurements would provide a valuable
resource with which to better understand the strengths and limitations
of this instrument and of similar low-dispersion, Cassegrain
spectropolarimeters. However, FORS1 data reduction has been carried
out by a number of different groups using a variety of reduction and
analysis techniques.  Our understanding of the instrument and our
data reduction techniques have both improved over time. A full re-analysis
of FORS1 archive data using a consistent and fully documented algorithm
would optimise the accuracy and usefulness of a catalogue of field
measurements.
}
{ 
Based on the ESO FORS pipeline, we have developed a semi-automatic
procedure for magnetic field determinations, which includes
self-consistent checks for field detection reliability.  We have
applied our procedure to the full content of circular
spectropolarimetric measurements of the FORS1 archive.
}
{
We have produced a catalogue of spectro-polarimetric observations and
magnetic field measurements for $\sim 1400$ observations of $\sim 850$ different
objects. The spectral type of each object has been accurately
classified. We have also been able to test different methods for data
reduction is a systematic way. The resulting catalogue has been used
to produce an estimator for an upper limit to the uncertainty in a
field strength measurement of an early type star as a function of the
signal-to-noise ratio of the observation. 
}
{
While FORS1 is not necessarily an optimal instrument for the discovery
of weak magnetic fields, it is very useful for the systematic study of
larger fields, such as those found in Ap/Bp stars and in white dwarfs.
}
\maketitle
\section{Introduction}
During a full decade of operations, the FORS1 instrument on the ESO
Very Large Telescope collected a large number of magnetic field
measurements of various kinds of stars. Together with the ESPaDOnS
instrument on the Canada-France-Hawaii Telescope, and the MuSiCoS and
NARVAL instruments on the 2\,m Telescope Bernard Lyot of the
Pic-du-Midi Observatory, FORS1 has been one of the workhorse
instruments for observational studies of stellar magnetism.

Most, if not all, FORS1 field measurements have been published in the
literature in dozens of different articles. Gathering them in a
general catalogue would serve to obtain an overview (even though
biased at the target selection phase) of the incidence of the magnetic
fields in various kinds of stars. However, a catalogue compiled using
published material would suffer from the lack of homogeneity in the
way data have been treated. Furthermore, over time, new ideas for data
reduction and quality checks have improved the reliability of FORS1
magnetic measurements, which calls for a revision of earlier data. We
also note that the literature of FORS magnetic field measurements
includes a certain number of controversial detections. These problems
have been thoroughly discussed by \citet{Bagetal12} and
\citet{Bagetal13}, and a discussion on the quality of the
non-controversial FORS1 measurements of magnetic Ap stars was
presented by \citet{Lanetal14}.

Here we publish our full catalogue of FORS1 measurements and explore
experimental relationships between signal-to-noise (S/N) ratios and
error bars achieved in stars with different spectral characteristics.

\section{Instrument and instrument settings}
FORS1 is a multi-purpose instrument equipped with polarimetric optics
capable of performing imaging and low-resolution spectroscopy in the
optical. It was attached to the Cassegrain focus of one of the 8\,m
units of the ESO Very Large Telescope (VLT) at the Paranal Observatory
from the beginning of operations in 1999 until instrument
decommissioning in March 2009.  The instrument is described in
\citet{AppRup92} and \citet{Appetal98}.

\subsection{Polarimetric optics}
The polarimetric optics of FORS1 are arranged according to the optical
design described by \citet{App67}. These components are embedded in
the overall optical train of the low-dispersion spectrograph for
spectropolarimetric observations as follows. The Cassegrain focal
plane of the telescope coincides with a mask containing 18 parallel
sets of positionable slit jaws, which in simple spectroscopy allow
multi-object spectroscopy of up to 18 objects simultaneously. For
spectropolarimetry every second pair of slit jaws is masked to prevent
beam overlapping in the camera \citep[following the scheme proposed
  by][]{Scaretal83}, so up to nine slits can be used at once. For most
observations only a single slit was used, normally (but not always)
centred on the optical axis of the telescope and of the spectrograph
collimator (``fast'' mode), but a number of spectropolarimetric
observations using the multi-slit capability were carried out for
studies of clustered objects (e.g. stars in an open cluster; ``fims''
mode). The slits are 22\arcsec\ long and can be adjusted to an
arbitrary width.

The slit plane is followed by a dioptric collimator consisting of four
UV-transmitting lenses, which takes each diverging beam from the
slit plane and converts it into a parallel beam; the
collimated beams from different slits have slightly different
axes. For spectropolarimetry, these collimated beams are then passed
through a rotatable super-achromatic quarter- or half-wave plate,
followed by a beam-splitting Wollaston prism which produces two
slightly diverging beams that have been divided into two orthogonal
linear polarisation states. Each beam pair is analysed into polarisation states
parallel to and perpendicular to the plane of the beam divergence
produced by the Wollaston prism.

Following the polarimetric optics, the beams pass through a grism, and
possibly an order-sorting filter, which disperses each beam into a
spectrum. This is followed by a camera lens system (four lenses) that
images the dispersed light from each polarised beam into a spectrum
along one axis of the CCD detector. The two dispersed beams from each
single slit are imaged on neighbouring CCD rows (in the case of
multi-slit observations, the various pairs of beams are arranged
parallel to one another on the detector). Spectropolarimetry is
accomplished (in principle) by comparing the two beams from each
single slit to form sum and difference spectra, from which a
polarisation Stokes parameter can be deduced.

\subsection{CCD and CCD readout}
Two detectors have been used in the FORS1 instrument: a 2k$\times$2k
SITE CCD (from the beginning of operations to end of February 2007),
and a mosaic composed of two 2k $\times$ 4k MIT CCDs with a pixel size
of $15 \times 15\,\mu$m (from March 2007 until FORS1 decommissining in
March 2009). The upgrade to the MIT CCD was described by \citet{Szeetal07}.

The older SITE CCD had a pixel scale of 0.20\arcsec. For most of the
observations obtained with it, the readout mode was set in ``low
gain'' (to minimise the ADU count, and the risk of saturation of the
ADC, at typically 2.8\,e$^-$ per ADU\footnote{The conversion
  from ADU to electron is recorded in the fits-header keyword {\tt
    DET.OUT1.CONAD}. However, in the QC1 database, the same quantity
  is called gain, while the QC1 entry {\tt CONAD} gives the number of
  ADU per electron. Conversely, the fits-header keyword {\tt
    DET.OUT1.GAIN} represents the conversion factor from electrons to ADUs,
  but corresponds to the entry {\tt CONAD} in the QC1 database\\
{\tt http://www.eso.org/observing/dfo/quality/}.}),
and with a window of 400 or 500 pixel rows centred about the spectrum,
to minimise CCD readout overheads, which represent a consistent
fraction of the total overhead time necessary to achieve high
S/N ratio spectropolarimetric measurements.

The MIT detector, composed of two chips, had a 0.125\arcsec\ pixel scale,
although in many observations a $2\times2$ rebinning
was adopted for the readout. The quantum efficiency of the MIT CCD in
the blue was higher than that of the SITE CCD, but the MIT CCD
suffered from heavy fringing in the red.  One of the advantages of the
MIT CCD compared to the SITE CCD was its better cosmetic character.
Figure~1 shows the raw image of a spectropolarimetric frame obtained 
in fast-mode with the MIT CCD. An internal reflection due to the
Longitudinal Atmospheric Dispersion Corrector \citep[LADC;][]{Avetal97},
visible in the blue edge of the CCD, has affected many observations. 

\begin{figure*}
\scalebox{0.255}{
\includegraphics*{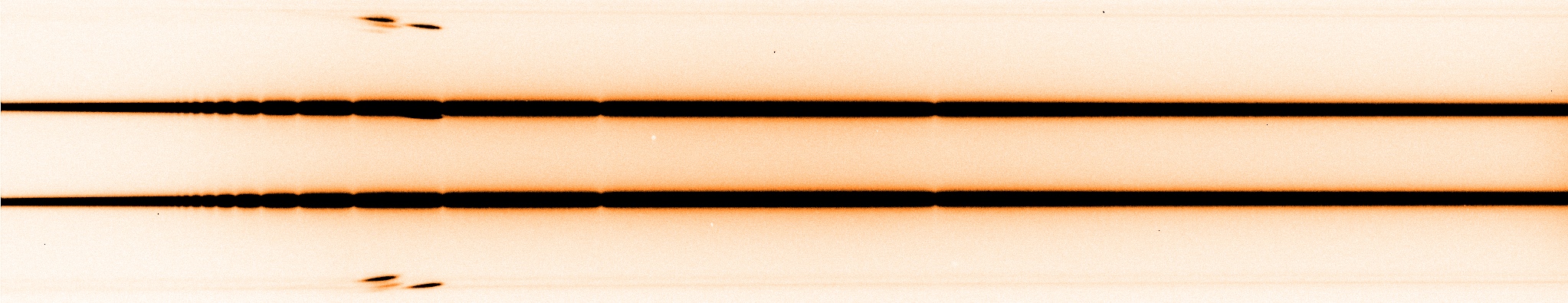}}
\caption{\label{Fig_CCD} Raw image of a polarisation spectrum obtained 
with the MIT CCD. On the blue (left) side reflections from the LADC are visible.}
\end{figure*}
Most of the observations were obtained in fast-mode, while only a
fraction of the observations were obtained in multi-object
mode, in which up to nine polarised spectra were obtained with the same
frame series. No windowing option was offered for the operations with
the MIT CCD, but its typical readout time was comparable to the readout time of
the SITE CCD when windowed to 4-500 pixel rows.
 
\subsection{Grisms and slit width}
In order of frequency of usage, most of the observations were carried
out with grisms 600B ($\sim 1000$ observations), 1200g ($\sim 100$),
1200B ($\sim 100$), 600R ($\sim 150$), and only very rarely with grism
300V ($\sim 25$) and 300I (2) The slit width was generally set to
0.4\arcsec\ or 0.5\arcsec, and very rarely $> 1\arcsec$. The
systematic use of narrow slits suggests that users wanted to have a
high spectral resolution and did not care much about slit losses.

Table~\ref{Tab_Instrument_Settings}, obtained from the
various editions of the FORS User manual, summarises the
characteristics of these grisms. We note that grism 1200\,g was often
used setting the slit close to the right edge of the instrument field
of view.  For that special setting, the observed wavelength interval
was often offset to the blue to include more Balmer lines than in the
configuration with the slit at the centre of the field of view. 
Grism 600R was used together with order separation filter GG\,435, while
600B, 1200B and 1200g were always used with no filter.

\begin{table}
\caption{\label{Tab_Instrument_Settings} 
Summary of the characteristic of the grisms+CCD most commonly employed for magnetic
field measurements. }
\begin{tabular}{rlccrr}
\hline \hline
Grism  & CCD   & Wavelength &    dispersion  &spectral   \\
       &       & range (\AA)& (\AA\ px$^{-1}$)&res. (1\arcsec)\\
\hline
 600B  & SITE & 3470--5900       &    1.20  &  780  \\
 600B  & MIT  & 3300--6210       &    0.70  &  800  \\
1200B  & SITE & 3800--4960       &    0.61  & 1420 \\
1200B  & MIT  & 3660--5110       &    0.43  & 1420 \\
1200g  & SITE & 4290--5470       &    0.58  & 1400 \\
 600R  & SITE & 5250--7420       &    1.08  & 1160 \\
\hline
\end{tabular}
\end{table}

\subsection{Observing strategy}
Most of the observations were obtained by setting the retarder waveplate
at two position angles relative to the principal plane of the
Wollaston prism, and obtaining multiple exposures for the purpose
of maximising the S/N ratio and allowing the computation of the
null profiles. The most typical observing sequence was
$-45\degr$, $+45\degr$, $+45\degr$, $-45\degr$,
$-45\degr$, $+45\degr$, $+45\degr$, $-45\degr$. This beam-swapping
technique allows one to minimise instrumental effects as explicitily
suggested in the FORS1/2 manual, and thoroughly discussed, e.g. by
\citet{Bagetal09}. \citet{Bagetal13} have argued that swapping between
only two positions of the retarder waveplate may lead to
more accurate results than cycling through all four positions in
quadrature (i.e.\ $-45\degr$, 45\degr, 135\degr, 225\degr) because
the latter sequence is more likely to introduce small instrumental
wavelength offsets between different exposures.
 
\section{Data reduction}
In this Section we give a detailed description of how we have
organised the archive data, and how we have treated them to measure
the circular polarisation. We will adopt the same formalism used in
\citet{Bagetal09}, i.e.\ \fo\ and \fe\ are the fluxes in the parallel
and in the perpendicular beam, respectively, $\pv = V/I$ is the
circular polarisation normalised to the intensity, and \nv\ is the
null profile (also normalised to $I$), a quantity that was introduced
by \citet{Donetal97}, and, as described by \citet{Bagetal09}, is
representative of the noise of \pv.

We have always obtained \pv\ profiles from a series of one or more
pairs of exposures. Each pair of exposures is composed of two frames
obtained with the retarder waveplate at postition angles separated by
90\degr.  In Sect.~\ref{Sect_Orga} we
explain the criteria followed to associate the frames retrieved from
the archive in series of polarimetric measurements, which in fact may
occasionally differ from the original plans of the observers.

For most of the observing series, it was also possible to calculate
the null profile. For those cases in which the number of pairs of
exposures $N$ was odd and $\ge 3$, the null profile was obtained
omitting the last pair of exposures. Obviously, with just one
pair of exposure, the null profile was not calculated.

\subsection{Organising frames}\label{Sect_Orga}
\subsubsection{Scientific frames}
As a first step we downloaded from the archive all frames obtained in
spectropolarimetric mode with the quarter wave retarder in the
optical beam. Then we grouped individual frames according to target
pointing and observing night.  Target identification was obtained via
cross-correlation between {\tt RA} and {\tt DEC} keywords and SIMBAD
catalogue, although we note that the fits-header keyword {\tt
  OBS.TARG.NAME} , which is set manually by the observer, turned out
to be sufficiently meaningful to identify the observed target in all
but a very few cases. Occasionally, the RA and DEC of a target with the same
{\tt OBS.TARG.NAME} slightly changed within a consecutive series of
exposures. We automatically ascribed a change of RA and DEC within
0.5\arcsec\ as due to a change of the guiding star; for larger offsets
we visually inspected the Stokes $I$ profile to check whether the
observations were in fact pointing to distinct components of a visual
multiple system. In the (rare) cases in which the same target was
acquired twice or more times during the same night after an interval of time
longer than 1\,h, the observations were split and treated as
independent field measurements. Most of the observation groups finally
included \textit{at least} two pairs of exposures, each pair with the
retarder waveplate at position angles $+45\degr$ and $-45\degr$. Some
observing sets included an odd number of exposures. In many cases,
this was because a short test exposure was obtained prior executing a
long series, with the aim of deciding on the exposure time. These
short exposures were then discarded.  Sets including only one exposure
were discarded.

The archive includes a few long time series of exposures that were
performed within the same night on rapidly rotating or pulsating
stars, and that were aimed at monitoring the target during its
rotation or pulsation cycle. Example of these cases include the roAp
stars observed within programme ID 69.D-0210 and 270.D-5023 (see
Table~\ref{Tab_IDs}), or the cataclismic variables II~Peg and V426~Oph
observed with programme ID 079.D-0697 and 081.D-0670.  In all
  these cases we had the choice whether to report the field values e.g. for
  each pair of frames, or to measure the
  field from the $I$ and \pv\ profiles obtained adding up all individual
  frames. For simplicity, we decided to adopt the latter approach. The
  interpretation of these field measurement has to be given case
  by case. For instance, since in roAp stars there is no evidence of
  a variability of the magnetic field with stellar pulsation, the
  value averaged over several pulsation cycles is still a meaningful
  estimate of the actual star's longitudinal field at a given rotation
  phase. If a time series extends over an interval of time that represents
  a non negligible fraction of the star's rotation cycle, then the averaged
  measurement may not be representative of the actual field. Long
time series may be identified by the of frames
used for field identification, which is an entry of our catalogue
(see Sect.~\ref{Sect_Entries}).

Any pair of frames where at least one beam in one exposure had an ADU
count $\ge 64\,000$ in at least 20 pixels was discarded as
saturated. Exceptions to this rule were applied when all pairs of
frames of a given series would be discarded, in which case we
rescued those spectral regions that were not saturated. A second
exception to this rule applies to the observations obtained in the
context of the observing programme 073.D-0464. The CCD gain had been
set to a very high value (3.5 ADU/e$^-$), with the consequence that
the CCD reached the full well capacity before ADC saturation. For all
frames obtained with that CONAD value we set the threshold for
saturation to 40,000\,ADUs instead of 64\,000\,ADUs.

\subsubsection{Calibration frames}
For each set of observations, we retrieved from the archive the
corresponding calibration frames, which included at least five bias
frames, one arc lamp, and one flatfield, although for each set, we
generally used five flatfield frames.  Most calibration frames were
obtained the morning after the night in which the scientific frames
were obtained. Occasionally, wavelength calibration frames were in
fact obtained one or two days later or earlier than science data, and
very rarely up to one or two weeks later or earlier. Time gaps between
science data and calibration frames longer than one day were found
more frequently for flat field calibrations. The reason is that
acquiring high S/N ratio
flatfield calibrations in the blue with a narrow slit is very
time consuming, especially with the polarimetric optics in. Hence,
for operational reasons, sometimes flatfield calibrations had to be postponed by
one or more days. We note that flatfield frames were used by the
pipeline to identify the CCD regions occupied by spectra, but
scientific frames were not divided by the flatfield.

\subsection{Deriving the Stokes and null spectra}\label{Sect_Spectra}
Each individual pairs of raw data were ingested into the FORS pipeline
\citep{Izzetal10} to perform bias subtraction, 2D-wavelength mapping of
the frames, and flux extraction without flat-fielding the science data.
Crucial for the data reduction was to avoid alignement with the sky lines.
Many targets were very bright, and the adopted short-exposure times were
not long enough to obtain high S/N ratio sky lines. Furthermore,
only one useful line is present in most of the settings. We found that
the alignment of each frame to the sky lines would generate differential
shifts that would be eventually responsible for spurious signals in
the polarisation spectra \citep[see Fig.~1 of][]{Bagetal13}.

We did not use the final pipeline products but we combined the various
\fo\ and \fe\ fluxes output by the ESO pipeline with a dedicated FORTRAN routine, and we obtained
\pv\ and \nv\ profiles using the formulas of the difference methods
given in Eqs.~(A2) and (A7) of \citet{Bagetal09}, which for convenience
we reproduce below,
\begin{equation}
\begin{array}{rcl}
\pv &=& {1 \over 2 N} \sum\limits_{j=1}^N \left[ 
\left(\frac{\fo - \fe}{\fo + \fe}\right)_{\alpha_j} - 
\left(\frac{\fo - \fe}{\fo + \fe}\right)_{\alpha_j + 90\degr}\right] \\[2mm]
\nv &=& {1 \over 2 N} \sum\limits_{j=1}^N (-1)^{(j-1)}\left[ 
\left(\frac{\fo - \fe}{\fo + \fe}\right)_{\alpha_j} - 
\left(\frac{\fo - \fe}{\fo + \fe}\right)_{\alpha_j + 90\degr}\right]\; ,\\
\end{array}
\label{EqVandN}
\end{equation}
where $\alpha_j$ belongs to the set $ \{-45^\circ$, $135^\circ \}$.
The reason for not using the final products of the pipeline was to
experiment with different algorithms. For instance, the rectification
that we use for \pv\ and \nz\ \citep[explained in Sect.~3.1
  of][]{Bagetal12} is carried out on the fluxes \fo\ and \fe.  This
rectification is occasionally needed for those cases in which we found
the \pv\ profile clearly offset from zero. This offset was found even
when no circular polarisation of the continumm was expected, for
instance in Herbig Ae/Be stars by \citet{Wadetal07}, and in several
other cases in the course of the present work. A possible explanation
is cross-talk from linear to circular polarisation, as discussed by
\citet{Bagetal09}. Obviously, cross-talk is expected to be a problem
only with observations of linearly polarised sources, and it is far
more significant for spectra acquired with a slitlet close to the edge
of the instrument field of view (as in some series obtained in
multi-object mode).

A slight but noticeable circular polarisation signal in the continuum
was also found in some of FORS data for sources that are {\it not}
linearly polarised, and that were observed in the centre of the field
of view. For these cases, we should probably rule out cross-talk as a
mechanism responsible for the observed continuum polarisation.  A
possible explanation could be that the ratio between the transmission
functions in the perpendicular beam $h^\perp$, and the transmission
function in the parallel beam, $h^\parallel$, is not constant as the
retarder waveplate rotates at the different position angles. In either
case (cross-talk from linear polarisation, or variability of the ratio
$h = h^\perp/h^\parallel$), the \pv\ profile should be rectified to
zero for a more accurate field determination. Inspection of the null
profile may help to discriminate between the two cases. If \pv\ is
offset from zero because of cross-talk from linear polarisation (or simply
because the source is intrinsically circularly polarised), the
null profile will still be oscillating about zero. If the \pv\ offset
is due to a non-constant ratio of the transmission functions, then
also the null profile will be offset from zero.

In this work, the \pv\ and \nv\ profiles were rectified to zero
as explained in Sect.~3.1 of \citet{Bagetal12}.

\begin{figure}
\scalebox{0.37}{
\includegraphics*{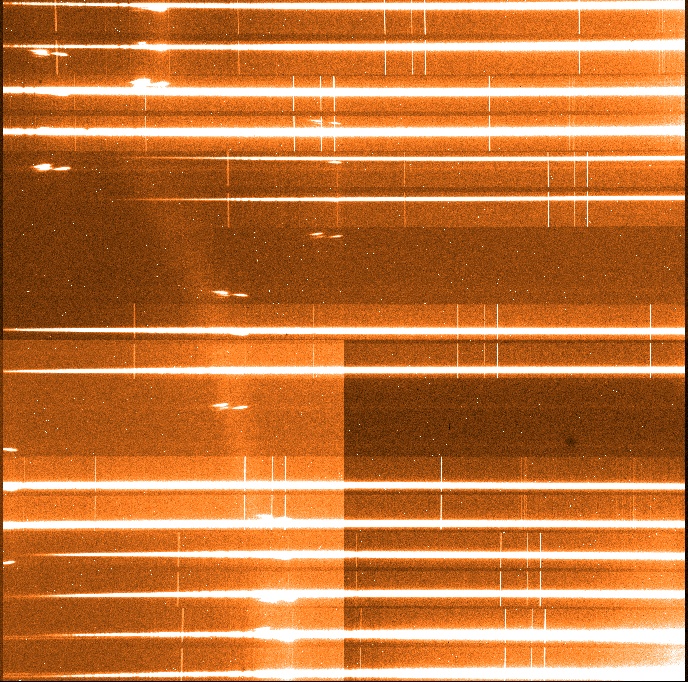}}
\caption{\label{Fig_Multi} Raw image of polarisation spectra obtained 
with the SITE CCD on 2003-02-09. Seven out of nine slitlets are on stars member
of an open cluster. The various reflections (presumably from the LADC) hamper
the automatic extraction and recombination of the beams by the pipeline.}
\end{figure}

An important difference concerns the treatment of data obtained
in multi-objects spectropolarimetric mode. In most of the cases,
the ESO pipeline failed to correctly associate the beams, probably
due to the presence of strong reflections in the frames. 
Figure~\ref{Fig_Multi} shows an example of a frame obtained in 
multi-object mode, with seven slits centred on a target, and two
slits closed. Some of the data obtained in multi-object mode
(mostly those pertaining to a large open cluster survey) were
``rescued'' through manual data reduction 
(see Sect.~\ref{Sect_Comments}).

Finally, in less than 3\,\% of the observations obtained in fast mode, the
pipeline delivered results of lower quality than expected.  Most of
these cases were succesfully individually treated by performing
a data reduction with IRAF tasks \citep{Fosetal15}.

\subsection{Magnetic fields determinations}\label{Sect_Bz_Meas}
\begin{figure*}
\begin{center}
\scalebox{0.69}{
\includegraphics*[angle=270,trim={1.0cm  0.9cm 1.5cm 1cm},clip]{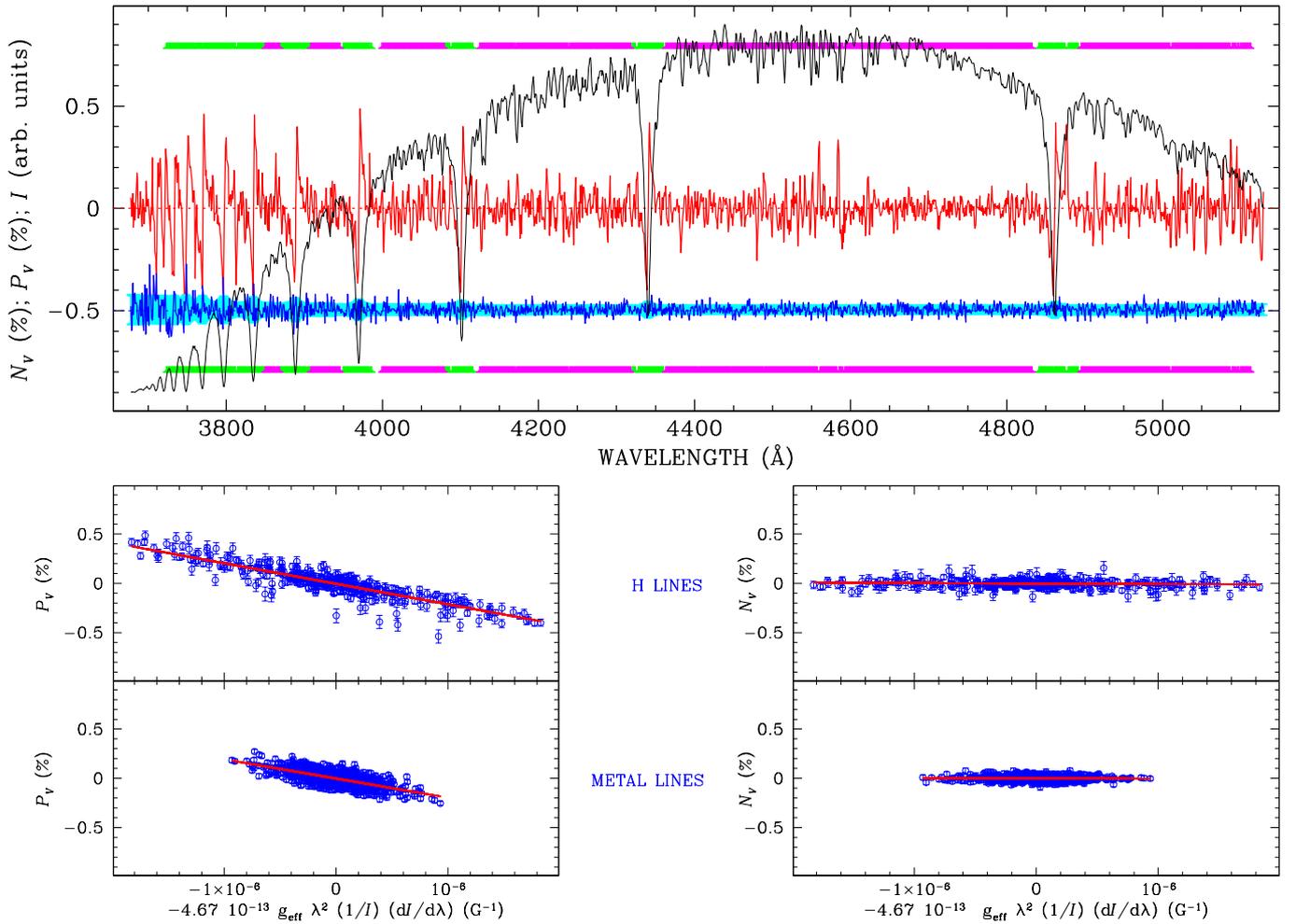}}
\end{center}
\caption{
\label{Fig_HD94660} An example of data reduction: the case of the Ap star HD\,94660.
In the upper panel,
the black solid line shows the intensity profile, the shape of which
is heavily affected by the transmission function of the atmosphere +
telescope optics + instrument. The red solid line is the \pv\ profile
(in \% units) and the blue solid line is the null profile offset by
$-0.5$\,\% for display purpose. Photon-noise error bars are centred
around -0.5\,\% and appear as a light blue background.
Spectral regions highlighted by green bars have been used to detemine
the \bz\ value from H Balmer lines, and the magenta bars highlight the
spectral regions used to estimate the magnetic field from metal
lines. The four bottom panels show the best-fit obtained by minimising
the $\chi^2$ expression of Eq.~(\ref{EqChiSquare}) using the
\pv\ profiles (left panels) and the \nv\ profiles (right panels) for H
Balmer lines and metal lines as described. The field values 
($\bz \sim -2000$\,G and $\nz \sim 0$\,G) are determined with a formal
accuracy of $\sim 40$\,G for Balmer lines and $\sim 25$\,G for metal lines.
}
\end{figure*}
FORS1 magnetic field measurements are obtained by exploiting the relationship
\begin{equation}
\frac{V}{I} = - g_\mathrm{eff} \ \cz \ \lambda^{2} \
                \frac{1}{I} \
                \frac{\mathrm{d}I}{\mathrm{d}\lambda} \
                \bz\;,
\label{EqBz}
\end{equation}
where $g_\mathrm{eff}$ is the effective Land\'{e} factor, and
\begin{equation}
\cz = \frac{e}{4 \pi m_\mathrm{e} c^2}
\ \ \ \ \ (\simeq 4.67 \times 10^{-13}\,\AA^{-1}\ \mathrm{G}^{-1}) \; ,
\end{equation}
where $e$ is the electron charge, $m_\mathrm{e}$ the electron mass,
$c$ the speed of light. We have adopted $g_{\mathrm{eff}} = 1.00$ for
the H lines, and 1.25 as an average for the metal lines.
\citet{Bagetal02} proposed to use a least-squares technique to derive
the longitudinal field via Eq.~(\ref{EqBz}), by minimising the
expression
\begin{equation}
\chi^2 = \sum_i \frac{(y_i - \bz\,x_i - b)^2}{\sigma^2_i}\; ,
\label{EqChiSquare}
\end{equation}
where, for each spectral point $i$, $y_i = \pv(\lambda_i)$, $x_i =
-g_\mathrm{eff} \cz \lambda^2_i (1/I_i\ \times
\mathrm{d}I/\mathrm{d}\lambda)_i$, and $b$ is a constant introduced to
account for possible spurious polarisation in the continuum. The
limitation of this method is that the spurious polarisation is
assumed to be constant in wavelength, which in fact may not be
true. The use of profiles rectified as explained in the previous
Section probably makes the introduction of the constant $b$ redundant.
The numerical evaluation of the quantity $1/I_i\ \times
(\mathrm{d}I/\mathrm{d}\lambda)_i$, which appears in the definition of
the term $x_i$, was obtained as
\begin{equation}
\frac{1}{I_i} \left(\frac{\mathrm{d}I}{\mathrm{d}\lambda}\right)_{\lambda = \lambda_{i}} =
\frac{1}{\mathcal{N}_{i}}
\frac{\mathcal{N}_{i+1} - \mathcal{N}_{i-1}}{\lambda_{i+1} - \lambda_{i-1}} \; ,
\label{EqDerivativeOne}
\end{equation}
where $\mathcal{N}_i$ is the photon count at wavelength
$\lambda_i$. 
\bz\ is calculated on points selected
after visual inspection either as pertaining to H Balmer lines or to
He and metal lines. 

We systematically avoided emission lines and spectral lines clearly
affected by non-photon noise (e.g.\ cosmic rays) and we generally
avoided using spectral regions judged featureless by means of a 
visual inspection.

We also obtained null field \nz\ in the same way as \bz, but using the
null profiles \nv\ instead of \pv.

Figure~\ref{Fig_HD94660} illustrates in detail how field is 
estimated from \pv\ and \nv\ profiles.
\begin{figure*}
\begin{center}
\scalebox{1.1}{
\includegraphics*[trim={2cm 7.9cm 2cm 6.3cm},clip]{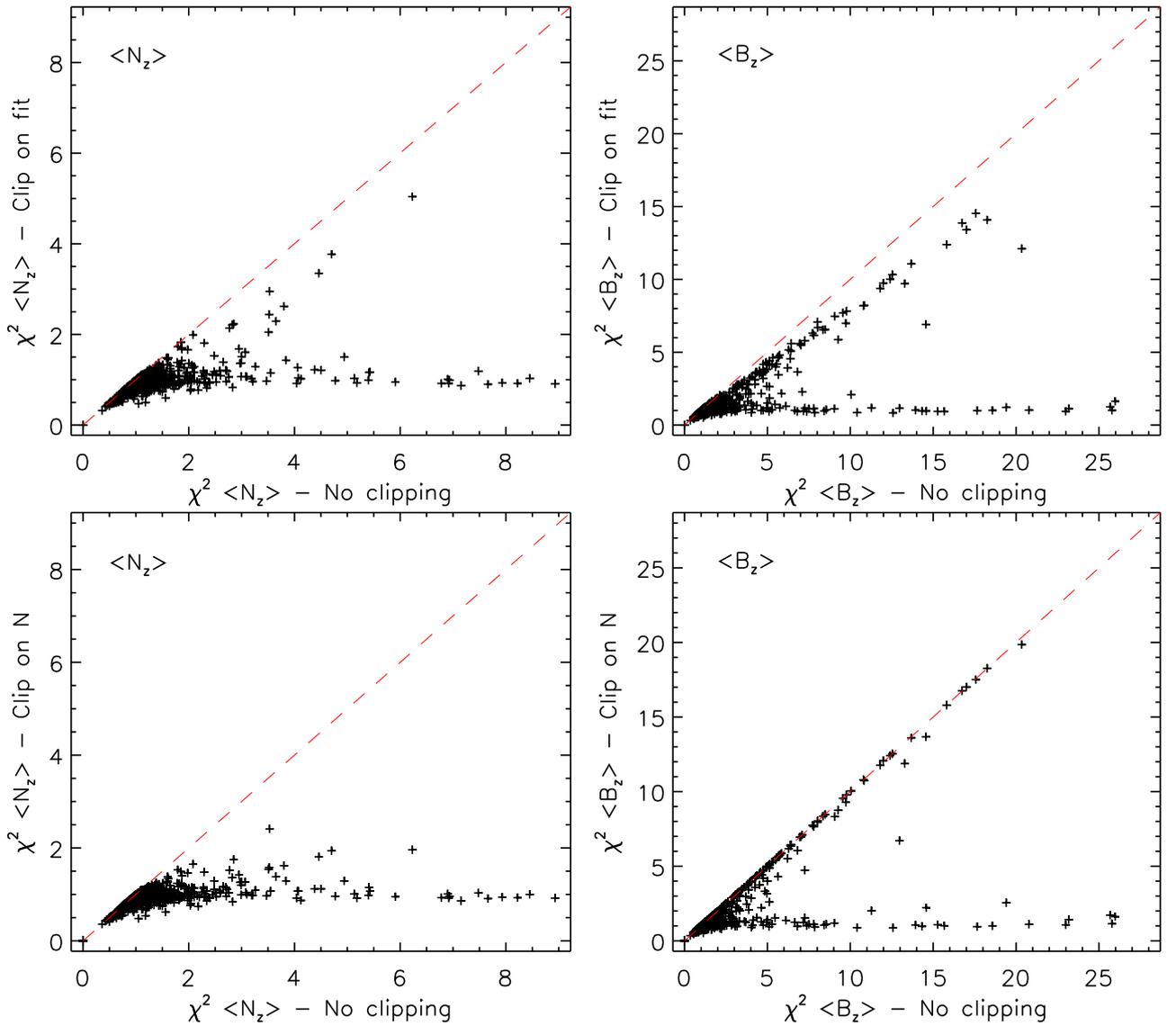}}
\end{center}
\caption{\label{Fig_Chi2} The reduced $\chi^2$ adopting two different
clipping algorithms for field estimates, versus the reduced $\chi^2$ obtained 
without adopting any clipping algorithm.
}
\end{figure*}

\subsubsection{Clipping algorithms}
It is possible to improve the quality of the field determinations by
applying one or more clipping algorithms. For instance, we can reject
all specral points for which the rectified null value exceeds
3-$\sigma$ in absolute value \citep[as proposed by][we call this
  method $N$-clipping]{Bagetal12}, or we can rejects all \pv\ and
\nv\ points that are more than 3\,$\sigma$ away from the interpolating
line $y = \bz x + b$ (we will call this method fit-clipping).

Another possible way to improve the precision measurement is to
reject all points for which $|x|$ is greater than a certain threshold,
as \citet{Lanetal12b} did. For instance, 
a point with $\vert x\vert > 10^{-6}$\,G$^{-1}$ in the spectrum of a
white dwarf is probably due to a cosmic ray rather than to a real
sharp spectral line, hence the motivation for this type of clipping.

Figure~\ref{Fig_Chi2} shows the effects of some of these algorithms on
the final error bars and on the reduced $\chi^2$, and makes the effect of their use evident.

We noticed that in several cases, a clipping on deviant \nv\ points
would improve the \nz\ best-fit, without having a significant impact
on the quality of the fit used to determine \bz. Conversely, the
clipping on the best-fit was found to be more efficient.  The reason
is that spikes in \pv\ due to instrumental instabilities, for example, do not
necessarily also appear in the null profiles, and vice versa
\citep[for a detailed discussion, see][]{Bagetal12}. We finally decided
to implement a $\sigma$\,clipping on the best-fit, adopting the
following specific algorithm \citep{Bagetal06}. As a first step, a best-fit is obtained
by minimising the expression of the $\chi^2$ given by
Eq.~(\ref{EqChiSquare}), considering all (pre-selected) spectral
points. Then we calculate the median and the median absolute deviation
(MAD) of the distances weighted by the photon-noise error between
\pv\ (or \nv) values and the best-fit. We then reject the \pv\ (or \nv) points
for which the weighted distance from the best-fit is $> 3 \times
1.48$\,MAD. The procedure is iterated until no points are rejected,
but from the second iteration on we reject the points that have
distance from the best fit-larger than three times the reduced
$\chi^2$ value.

It is important to recognise that for a given dataset
it is not possible to associate a uniquely defined longitudinal field
estimate.  \citet{Bagetal12} have thoroughly discussed how two equally
reasonable data reduction procedures produce (slightly) different
results. In some cases one may be able to decide that one procedure
may be more appropriate than another one, but in most cases we are left
with a certain degree of arbitrariness.  Among steps that may affect
the final results one should consider whether data are flatfielded or
not, which method is adopted to extract spectra (average extraction or
optimal extraction), if and how Stokes profiles are rebinned and/or
rectified, how the derivative is calculated, if and how data are
clipped, which spectral regions are used for the field determination,
and which effective Land\'e factor is adopted (the latter choice does
not change the relative error measurement). It is not surprising
therefore that from the same dataset, different field values are
obtained by different authors, or even by the same authors in
different epochs. In this respect, data reduction should be somehow
considered as a source of noise that adds to photon-noise and
instrument instabilities.

\subsection{Error bars}\label{Sect_ErrorBars}
Error bars of \bz\ and \nz\ are calculated using Eqs.~(10) and (11) of
\citet{Bagetal12}. Briefly, they are calculated by propagating the
photon-noise of the fluxes, and then multiplied by the square root of
the reduced $\chi^2$. When a field is detected,
the reduced $\chi^2$ associated with the \bz\ estimate tends to be
higher than the reduced $\chi^2$ associated with the \nz\ estimate. As
a consequence of the way they are calculated, the \bz\ error bars are
also systematically higher than \nz\ error bars. Since our error bars
are proportional to the square-root of the reduced $\chi^2$, field
error bars are higher in strongly magnetic stars than in weak-field or
non-magnetic stars. This phenomenon is not surprising, but is simply a
natural consequence of the fact that Eq.~(\ref{EqBz}) is only an
approximation, and a lot of effects conspire to deviate from it, such
as line blending, breaking of the weak-field approximation in metal
lines, Lorentz forces and Stark broadening of the H lines. One could
even conclude that a situation where the reduced $\chi^2$ associated
with \pv\ is substantially higher than the $\chi^2$ associated with
\nv\ represents already {\it per se} an indication that a magnetic
field is present.
  
\section{Precision of field measurements versus spectral signal-to-noise ratio}
\begin{figure*}
\scalebox{0.45}{
\includegraphics*[trim={1cm 6.2cm 0cm 3.1cm},clip]{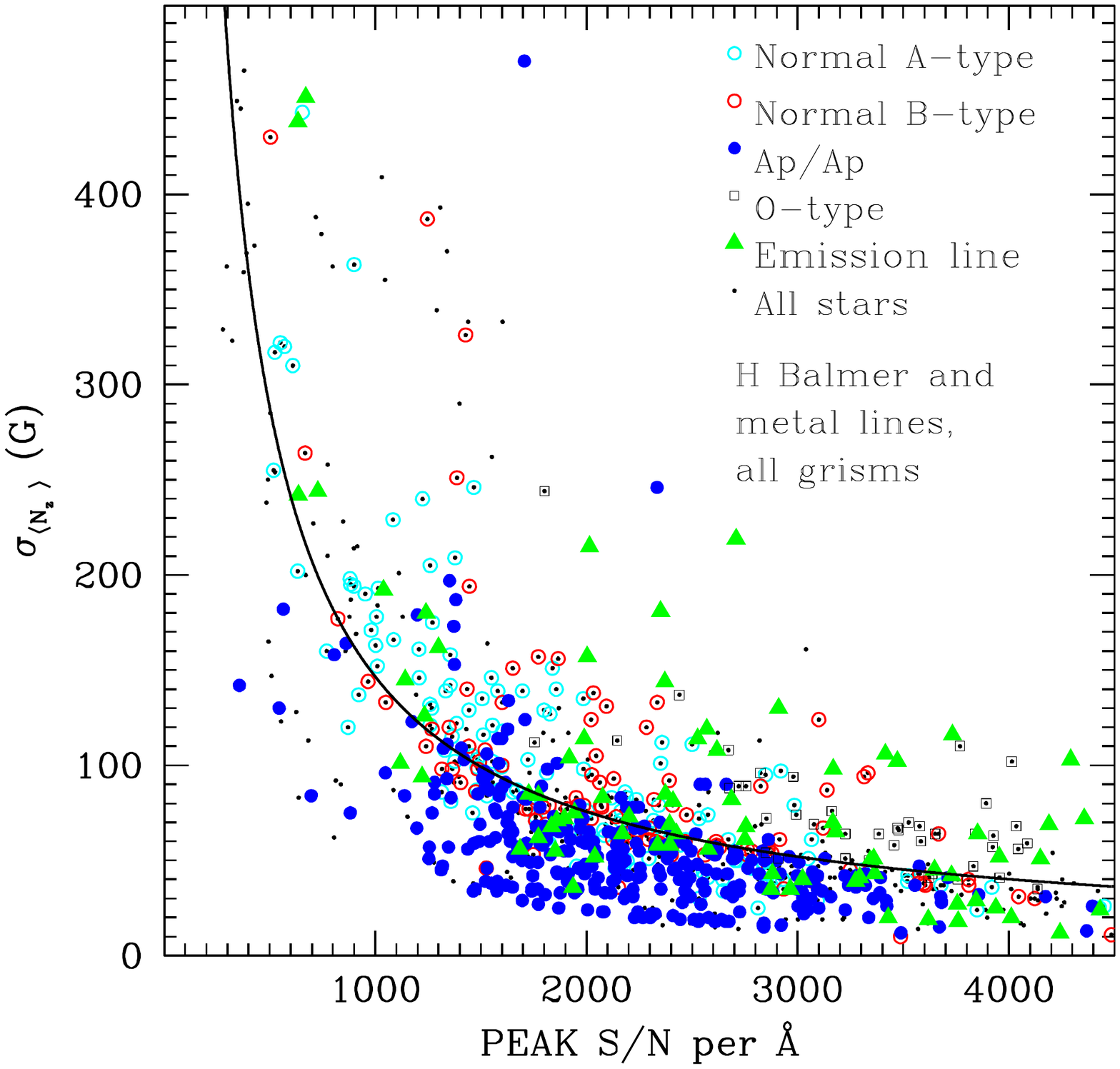}}
\scalebox{0.45}{
\includegraphics*[trim={1cm 6.2cm 0cm 3.1cm},clip]{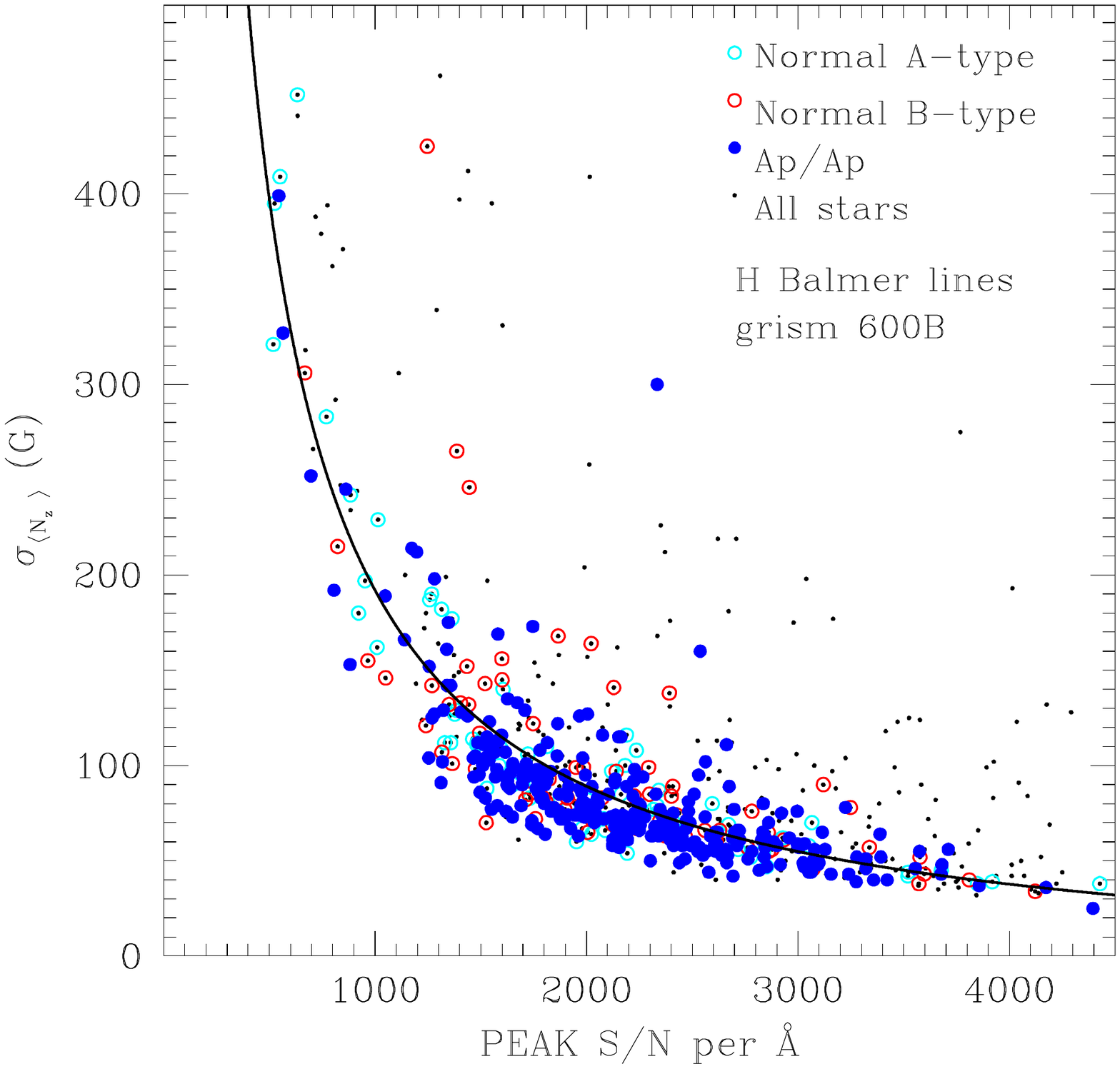}}
\vspace{0.2cm}

\noindent
\scalebox{0.45}{
\includegraphics*[trim={1cm 6.2cm 0cm 3.1cm},clip]{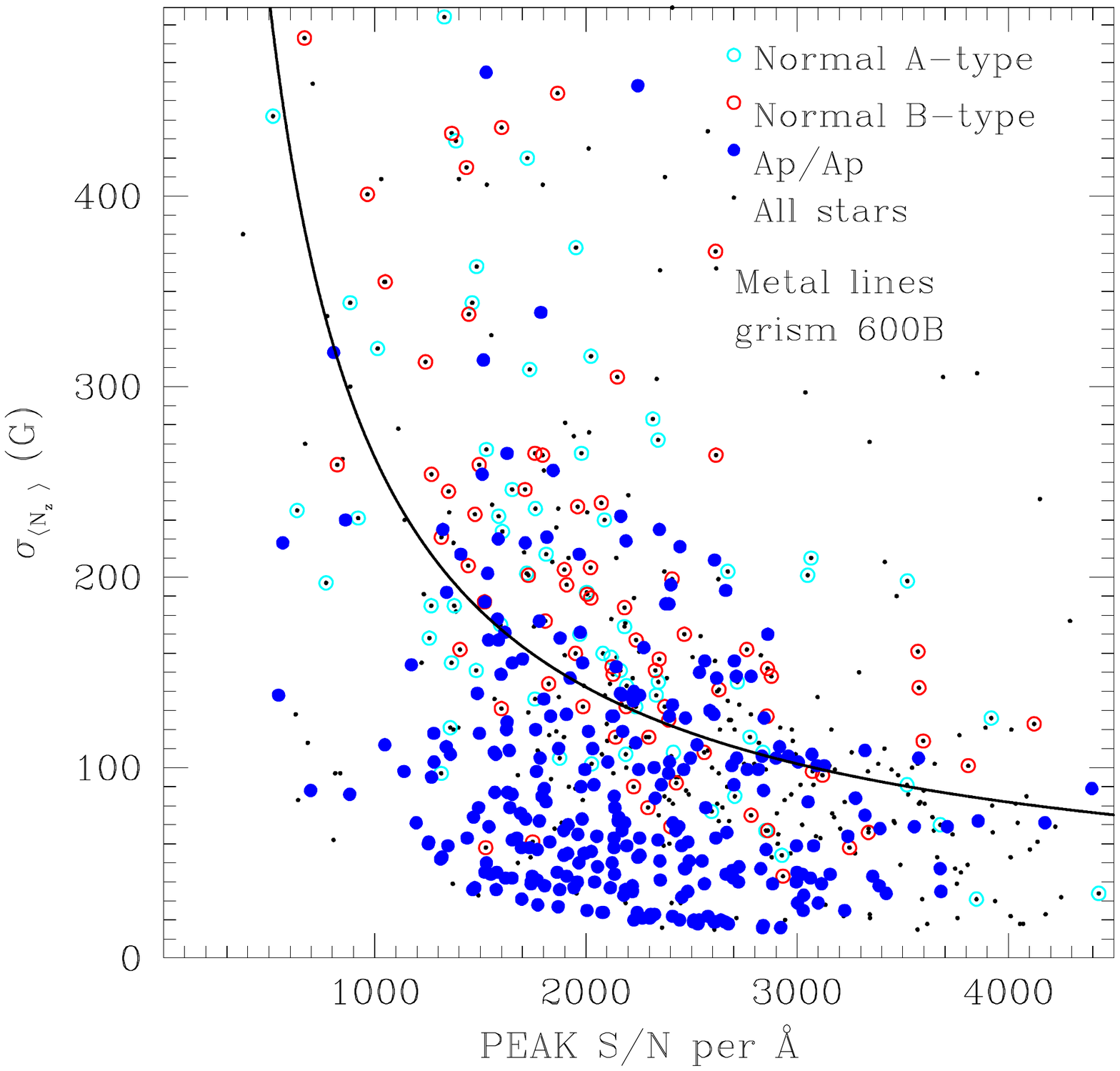}}
\scalebox{0.45}{
\includegraphics*[trim={1cm 6.2cm 0cm 3.1cm},clip]{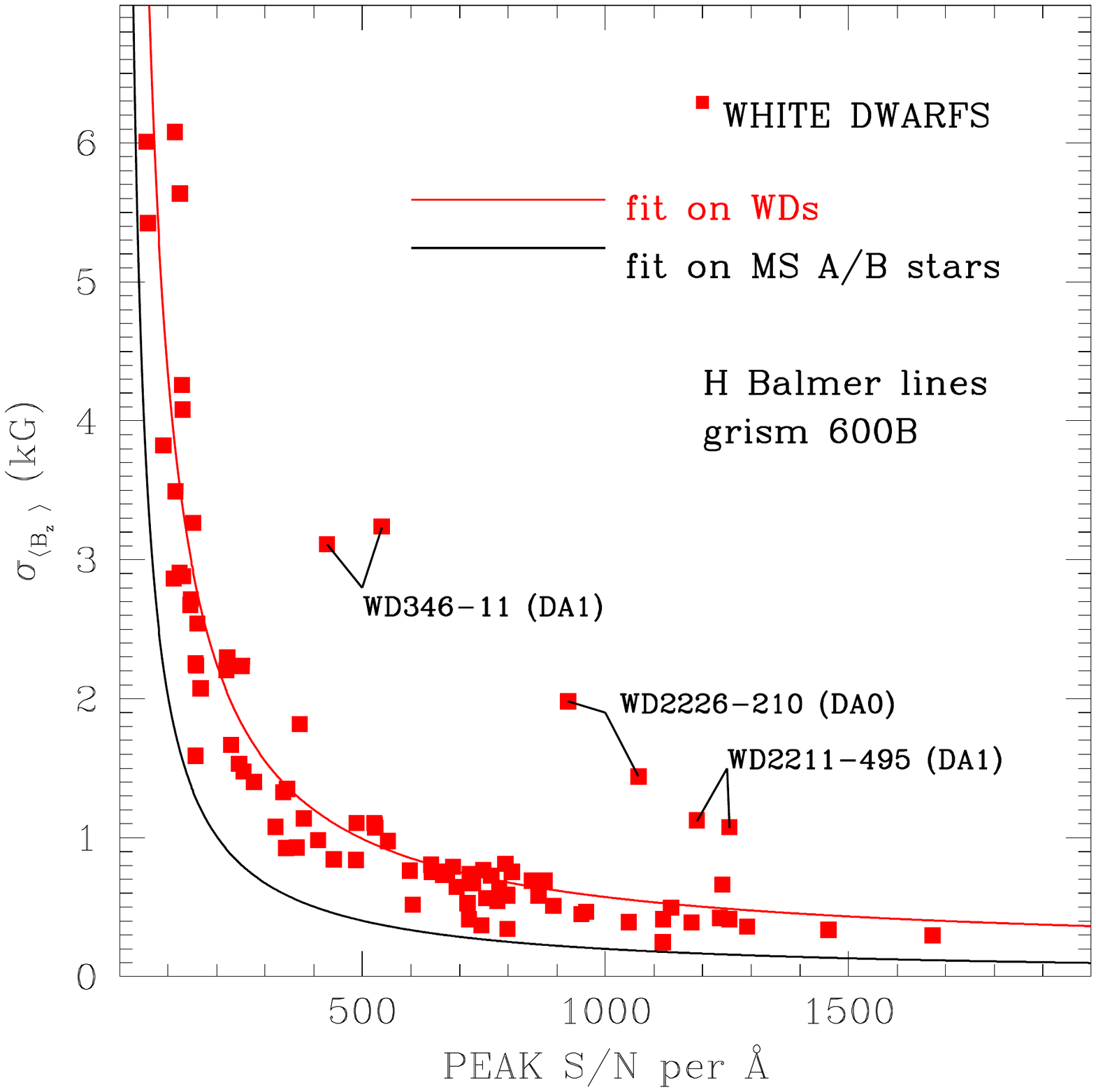}}
\caption{\label{Fig_SNR} Error bars versus
the peak S/N ratio for different kinds of stars. 
Left top panel: the error bars on the null field for all stars
calculated using both H Balmer and metal lines. 
Right top panel: same as left panel, but considering only main
sequence A- and B-type stars (including Ap/Ap stars) 
observed using grism 600B, and using H Balmer lines only.
Left bottom panel: same as right top panel, but using metal lines
only. 
Right bottom panel: the errors \sbz\ on the longitudinal field
measured from H Balmer lines only in white dwarfs; the outliers
are the hottest star.
}
\end{figure*}
Our catalogue may be used for a number of statistical studies, some of
them already discussed by \citet{Bagetal12}. 

An important question for planning observations and for evaluating
their success is the exent to which the final magnetic field
uncertainty of a measurement may be predicted from the expected S/N
ratio of the observation (as estimated for example using the exposure
time calculator, or measured after the observation). We here use the
catalogue to establish a quantitative link between field error bars
and spectral S/N ratio for stars of different spectral classes.

A global overview of the situation is shown in the upper left panel of
Figure~\ref{Fig_SNR}. This panel shows the error bar \snz\ of the null
field versus the peak S/N ratio per \AA\ measured in the spectrum for
most of the main spectral classes present in the catalogue, using the
value of \snz\ computed from the entire (useful) spectrum. The reason
to choose \snz\ rather than \sbz\ as abscissa is that, as discussed in
Sect.~\ref{Sect_ErrorBars}, in stars with strong magnetic fields,
\sbz\ may be much higher than \snz, without being representative of
the real detection threshold.  In this panel the data do show a broad
trend of decreasing \snz\ with increasing peak S/N ratio, and we
already see that to obtain measurements with $\snz \la 100$\,G,
it is necessary to obtain spectra with peak S/N ratio $\ga 10^3$.

The data also show a considerable scatter around the mean curve.  Part
of this scattering occurs because in some spectral classes a more
precise field measurement is possible than in others (for example,
field measurements are more precise in Ap/Bp stars with their rich
line spectra than in O-type stars with far fewer, weaker lines). Part
of this scatter seems to be intrinsic to each main spectral class.

The scatter around the mean curve may be better understood by
restricting the data used. In the upper right panel of
Fig.~\ref{Fig_SNR}, the uncertainty \snz\ is plotted again for normal
A and B stars and for Ap/Bp stars, but using only observations with
the 600B grism and only the Balmer line regions for the estimate of
\snz. We see that the dispersion around the mean curve (which here is
fitted to the data, with somewhat arbitary omission of some outliers)
is dramatically reduced. The tightness of the data around the mean
curve reflects the fact that the Balmer line strength and shape do not
change very much through this effective temperature range, and are not
very different for Ap/Bp stars than for normal stars.

In contrast, the left bottom panel shows \snz\ as determined using
only the metallic line regions and grism 600B. Here the scatter is
much larger than in the preceding panel, showing the diversity of the
metallic spectra in this sample. In this sample, the number of
available spectral lines, and their typical depth and breadth, vary
with spectral class and with projected rotational velocity, and some stars
have many points in the spectra with steep slopes d$I/$d$\lambda$,
while others have few or none. In turn this has a very strong effect
on the horizontal distribution of points in correlation diagrams such
as the lower panels of Fig.\ref{Fig_HD94660}, with the consequence that
for some stars the field is far more precisely determined from metal
lines than for others. At the extremes, \snz\ from metal lines may be
much less than half as large as that from the Balmer lines, or twice
as large. Thus the scatter in the upper left panel arises essentially
from the variations in the metallic line spectrum. For early-type
stars, a robust estimation of the field uncertainty expected from the
H lines as a function of S/N ratio may be obtained, but the
(potentially quite large) improvement in this basic \snz\ coming from
the metallic lines cannot even be estimated without knowing in some
detail the nature of the metallic line spectrum.

The right bottom panel of Fig.~\ref{Fig_SNR} illustrates the special
case of DA white dwarfs. Here we have plotted \sbz\ instead of \snz\
because with field measurements based only on completely unblended H
lines (hardly any DA white dwarfs show lines of any other chemical
element), all with the same Land\'{e} factor, no extra dispersion
enters the evaluation of \sbz\ compared to \snz. It is seen 
that field measurements of the DA white dwarfs follow a similar
relation to the Balmer line relationship for main sequence A and
B stars. However, for a given S/N ratio, the field uncertainties are
two or three times higher than for main sequence A and B stars.
This larger uncertainty is essentially
a consequence of the fact that the H lines of white dwarfs are several
times broader than those of  main sequence stars with similar temperature.
Consequently, the slopes d$I/$d$\lambda$ of various points in the
spectrum of a white dwarf are smaller than in a A or B-type star. 
Since we have essentially $\sbz \propto \sigma_V \, ({\rm d}I{\rm d}\lambda)^{-1}$
it is clear that for a given value of the S/N ratio, \sbz\ is larger 
in stars with small d$I/$d$\lambda$ than in stars with higher d$I/$d$\lambda$.
We also note that there are a
few distant outliers in the white dwarf panel, because
white dwarfs with very high or very low effective temperatures have
very shallow Balmer lines that provide a very poorly constrained slope
in Fig.~\ref{Fig_HD94660}.

Finally, we note that, with the exception only of the fast rotating
G-type star FK~Com, FORS1 was never used to observe cool
stars. Clearly, from the higher density spectra of cooler stars we
expect to measure fields with higher precision than in chemically
normal early-type stars (but higher precision does not
automatically translate into higher accuracy). Field measurements in
FK~Com have reached a precision of $\sim ~20$\,G for a S/N ratio
$~\sim 3500$, comparable to what is achievable from metal lines of Ap
stars.

\section{Stellar classification}\label{Sect_Classification}
\begin{table*}
\begin{center}
\caption{\label{Tab_Classes} Symbols used in classification entry to the
catalogue of FORS1 magnetic field measurements}
\begin{tabular}{ll}\hline 
\textbf{Symbol} & \textbf{Description}  \\ 
\hline
\multicolumn{2}{l}{{\bf First field:} evolution state} \\[1mm]
PM    & (probable or certain) pre-main-sequence star \\
MS    & main sequence star, MK luminosity class IV or V \\
GS    & giant star, MK luminosity class III \\
SG    & supergiant star, MK luminosity class I or II \\
SD    & (hot) subdwarf \\
CP    & central star of planetary nebula \\
WD    & white dwarf \\
??    & star of unknown evolutionary state \\[2mm]
\multicolumn{2}{l}{{\bf Second field:} temperature class} \\ [1mm]
A5, B9, etc.   & temperature (spectral) class according to MK or HD system \\
DA7, DB3, etc. & temperature class according to white dwarf system of \citet{Sioetal83} \\
CV    & cataclysmic variable system (nova, dwarf nova, etc.) \\[2mm]
\multicolumn{2}{l}{{\bf Third field:} distinctive characteristics} \\
AM    & metallic-line (Am) star \\
AP    & star of the Ap (peculiar A) spectroscopic class \\
BCEP  & early B-type $\beta$~Cep pulsating star \\
CSD   & star showing evidence of a circumstellar debris disk \\
CV    & cataclysmic variable system (nova, dwarf nova, etc.) \\
DSCT  & $\delta$~Sct star, a late A-type pulsating main sequence star \\
E     & presence of emission lines, especially in H$\alpha$ \\
EB    & eclipsing binary system \\
FKCOM & FK Com variable (rapidly rotating cool giant) \\
FLS   & flare star \\
FP    & O stars with the Of?p peculiarity \\
HES   & a star showing abnormally strong He lines for its effective temperature \\
HEW   & a star showing abnormally weak He lines for its effective temperature \\
HGMN  & a late B-type star showing the HgMn class of spectral peculiarities \\
HIPM  & a high proper motion star \\
LPS   & standard of linear polarisation \\
M     & a star in which a magnetic field has definitely been detected (not necessarily with FORS1) \\
NOV   & nova \\
P     & peculiar (often chemically peculiar) \\
ROAP  & rapidly oscillating Ap star (roAp star) \\
SB    & spectroscopic binary system \\
SPB   & slowly pulsating B star \\
V     & variable \\
XRB   & X-ray binary system \\
\hline
\end{tabular}
\end{center}
\end{table*}

The stars of the FORS1 archive cover a wide range of spectral types
and evolution stages. Many are in some way distinctive or even
peculiar. Here we provide a classification to facilitate the
identification of all FORS1 measurements of a particular type of star.

In order to maximise the usefulness of these classifications, we went
beyond the very inhomogeneous classes that would be obtained, for
example, by simply taking the stellar classifications (often on the MK or
Henry Draper systems) from the Simbad data archive. One reason for
doing this is that the normal spectral classes reported in catalogues
are also quite limited in their intent; most simply describe the
morphological class to which an observed classification spectrum
belongs. Spectral classification often lacks important information
about the nature or evolutionary state of the star, and usually does
not contain any information about such important characteristics as
pulsation properties.

Our classifications are intended to provide a brief summary of a
number of different kinds of information about each of the stars
observed. We include information about {\it a)} the general
evolutionary state (pre-main sequence, main sequence, giant, etc.);
{\it b)} the photospheric temperature; {\it c)} some specific
distinctive features of the star such as chemical peculiarity,
pulsation properties, presence of a disk, clear presence of a magnetic
field in the star, etc. The system we have adopted is to provide a
single entry for each observation in the general format:\\ 
(evolution state):(effective temperature):(feature 1).(feature
2). \dots (feature N).\\

The choice of classes for the first field (evolution) is
a fairly obvious extension of the MK luminosity classes, supplemented
by classes not covered by this system such as CP for the central star
of a planetary nebula. 

In the second field (effective temperature), we follow the MK classes
whenever these provide a useful description of temperature (A5, B0, or
simply A or B if the temperature class is rather uncertain). For white
dwarfs we use the temperature classification system described by
\citet{Sioetal83}, where the numeral following the spectral type DA,
DB, etc., is the quantity $50400/T_{\rm eff}$ rounded to an integer.
Sometimes we have simply provided a class which explains why we cannot
give a simple temperature (such as CV for a cataclysmic variable). In
the case of SB systems, we have usually given the temperature class of
the brightest member.

In the third field (features), we have included a very wide variety of
information, including chemical peculiarity (metallic-line Am
star, helium-strong star), binary nature, evidence of circumstellar
material (classical Be stars, shell stars), pulsation properties
($\beta$\,Cep stars), clear presence of a magnetic field, and/or other
miscellaneous information.

The origins of the information contained in these classifications are
very diverse. We have naturally made extensive use of the classes
provided by the Simbad database, or more specific catalogues such as
the revised Henry Draper catalogue of Houk and collaborators
\citep{HouSwi99}. The white dwarf catalogue of \citet{McCSio99} has
been consulted extensively. In addition, we have often found useful
information in the publications that have been based on the data
re-analysed here. A large number of such articles are cited by
\citet{Bagetal12}. 

For a few of the spectra in the catalogue, literature spectral
classifications appeared to be inconsistent with the observed $I$
spectra. Furthermore, two observing programmes measuring magnetic
fields of open cluster stars (68.D-0403 and 70.D-0352) made extensive
use of the fims multi-object observing mode in order to observe as
large a sample of stars as possible. In these multi-object
observations, stars other than the one or two explicitly chosen
targets were simply selected from the nearby field. Such stars often
have no published classifications at all, even if they have cluster
numbers.

For cases of dubious or missing spectral classification in the
catalogue, we tried to assign spectral classes based on the observed
intensity spectra. We did this by visually comparing each observed
spectrum with a grid of spectra of stars of known spectral type, also
taken from the catalogue. For B and A stars classification was usually
successful, using normal spectral classification criteria such as line
strength ratios of lines of H, He and Ca~K. We could also often use
the width of Balmer lines to assign evolutionary state. The precision
of our spectral classes is estimated to be about $\pm$~2 or 3 spectral
subclasses, probably about the same as the precision of classes taken
somewhat randomly from the literature. However, this procedure usually
failed, or produced only very imprecise classification for late type
stars, for which very few stars of known spectral type are available
in the catalogue.  We were also usually unsuccessful in assigning
spectral classes to stars for which only spectra around H$\alpha$ are
available in the catalogue, as this region lacks the required variety
of clear classification indicators.

The full list of symbols and abbreviations adopted for our
classifications are given in Table~\ref{Tab_Classes}. For example, a
main sequence A9V star that is a $\delta$ Sct variable and a member of
a spectroscopic binary system will be classified as MS:A9:DSCT.SB.

\section{Description of the catalogue}
The FORS1 archive of circular spectro-polarimetric data includes
about 1500 observing series, for a total of more than 
about 12\,000 scientific frames, obtained within the context
of 59 observing programmes, using more than 2\,000 hours 
of granted telescope time and about 340 hours of shutter
time.\footnote{The top right panel of Fig.~6 of \citet{Bagetal12}
shows that a large fraction of FORS1 magnetic field measurements were
obtained with very short exposure times, in some cases even less than
1\,s. As a consequence, the execution time of many observing programmes were
dominated by overheads.} The content of the
FORS1 archive is presented here in the form of three printed tables,
one catalogue online, and one database of intensity
spectra available at {\tt http://star.arm.ac.uk/FORS/}. In the following
we describe this material.

\subsection{List of the observing programmes}
\begin{table*}
\caption{\label{Tab_IDs} List of observing programmes carried out with FORS1 in circular spectropolarimetric mode.}
\begin{tabular}{lllrrr}
\hline
 PR.    ID   & PI      &TARGETS              &TIME& YEAR& GRISM\\
\hline\hline
             &         &                      &    &     &      \\
  60.A-9203  & SCIOPS  & (engineering ID)     &    &     &      \\
  60.A-9800  & SCIOPS  & (engineering ID)     &    &     &      \\
  65.H-0293  & Jordan  &WD LP 790             &  6h& 2000&  150I\\
  65.P-0701  & Wagner  &Blazars               &  2N& 2000&  150I\\
  66.D-0128  & Appenzeller& Polar EF Eri      &0.7N& 2000&  600B\\ 
  67.D-0306  & Bagnulo &WD\,1953--011 (monitoring) & 13h& 2001&  600R\\
  68.D-0403  & Bagnulo &Open cluster Ap stars &2N&2002&  600R\\
  69.D-0210  & Hubrig  &Time series roAp stars&1n&2002& 600B\\
 269.D-5044  & Hubrig  &Mini-survey of roAp stars    & 10h& 2002&  600B\\
  70.D-0259  & Jordan  &Weak fields in white dwarfs    & 24h& 2003&  600B\\
  70.D-0352  & Bagnulo &Open cluster Ap stars &3h+2N& 2003&  600B\\
 270.D-5023  & Kurtz   &Time series of roAp star HD 101065&4h&2003&  600B\\
  71.D-0308  & Hubrig  &Evolution of Ap stars in the field& 20h& 2003&  600B\\
 072.D-0119  & Marsh   &Polar ES Cet         &  1N& 2003&  300V\\
 072.C-0447  & Bagnulo &Herbig stars         & 3HN& 2004&  600B\\
 072.D-0089  & Jordan  &Planetary Nebulae    &  7h& 2003&  600B\\
 072.D-0290  & O'Toole &Hot subdwarfs       &  1N& 2004&  600B\\
 072.D-0377  & Hubrig  &Evolution of Ap stars in the field&30h & 2004&  600B\\
 272.C-5063  & Bagnulo &Herbig stars         &4.5h& 2004&  600B\\
 073.D-0322  & Reinsch &Zeeman tomography of WDs&2h+3N&2004&  300V\\
 073.D-0356  & Jordan  &Weak fields in White Dwarfs   & 24h& 2004&  600B\\
 073.D-0464  & Hubrig  &Evolution of Ap stars in the field& 30h& 2004&  600B\\
 073.D-0466  & Hubrig  &SLP B and Bp stars        & 30h& 2004&  600B\\
 073.D-0498  & Bagnulo &Open cluster Ap stars& 30h& 2004&  600B\\
 073.D-0516  & Bagnulo &Cool White Dwarfs              & 42h& 2004&  600B\\
 073.D-0736  & O'Brien &X-rays binary        &  6h& 2004&  300V\\
 274.D-5025 & Mason   &Nova V574 Pup        &7.1h& 2004&  300V\\
 272.D-5026  & Bagnulo &Open cluster Ap stars&4.5h& 2005&  600B\\
 074.C-0442  & Bagnulo &Herbig stars         &  3N& 2004&  600B\\
 074.C-0463  & Yudin   &Vega-like stars      &  8h& 2005& 1200g\\
 074.D-0488  & Bagnulo &Open cluster Ap stars&4h+2N& 2005&  600B\\
 075.D-0289  & Jordan  &Planetary nebulae    &  3N& 2005&  600B\\
 075.D-0295  & Briquet &Pulsating B-type stars&30h& 2005& 1200g\\
 075.D-0352  & O'Toole &Hot subdwarfs        & 19h& 2005&  600B\\
 075.D-0432  & Schnerr &O-type stars         & 17h& 2005&  600B\\
 075.D-0507  & Yudin   &Be-type stars        & 12h& 2005& 1200g\\
 076.D-0435  & Berdyugina&White dwarfs       &  1N& 2005&  600B\\
 077.D-0406  & Yudin   &Be-type stars        & 10h& 2006&  600B\\
 077.D-0556  & Schmitt &X-ray A-type stars   &  1N& 2006&  600B\\
 277.D-5034  & Greiner &Cataclysmic variable V504 Cen &  2h& 2006&  600B\\
 078.D-0140  & Briquet &Pulsating B-type stars&16h& 2007&  600B\\
 078.D-0330  & Hubrig  &Hanle effects in high-mass stars& 18h& 2007&  600R \\
 278.D-5056  & Briquet &$\theta$~Car         &  1h& 2007& 1200B\\
 079.D-0240  & Mathys  &roAp stars           & 28h& 2007&  600B\\
 079.D-0241  & Briquet &B-type stars         &  2N& 2007&  600B\\
 079.D-0549  &Karitskaya &Cyg X-1/HDE226868  &  7h& 2007& 1200B\\
 079.D-0697  &Jeffers  &II Peg\& V426 Oph    &  2N& 2007& 1200B\\
 279.D-5042  & Hubrig  & $\upsilon$ Sgr      &2.5h& 2007& 1200B\\
 080.D-0170  & Mathys  & HD75049             &6.2h& 2008&  600B\\
 080.D-0383  & McSwain &Be-type stars        &  2N& 2008&  600B\\
 080.D-0521  & Kawka   & White dwarfs        & 74h& 2008&  600B\\
 081.D-2005  & Barrera & HD\,182180          &  2h& 2008& 1200B\\
 280.D-5075  & Korhonen&G-type giant star FK Comae&7.5h&2008&600B\\
 380.D-0480  & Yudin   &Be-type star $\lambda$ Eri&2H&2007&1200B\\
 081.C-0410  & Cure    &Herbig stars         &  2N& 2008&  600B\\
 081.D-0670  & Jeffers & II Peg \& V426 Oph  &  2N& 2008& 1200B\\
 381.D-0138  &Karitskaya& Cyg X-1/HDE226868  & 10h& 2008& 1200B\\
 082.D-0342  &Kolenberg& RR Lyrae stars      &2.5N& 2008& 1200B\\
 082.D-0695  & Reinsch & Accreting white dwarfs &  3N& 2008&  300V\\
 082.D-0736  &Vornanen & White dwarfs        &  2N& 2008&  600B\\
 282.C-5041  &Hubrig   & Z CMa               & 2.7h&2008& 1200B\\ 
\hline
\end{tabular}

\end{table*}

Table~\ref{Tab_IDs} is the list of the IDs and PIs of the observing
programmes, and their general characteristics (i.e.\ the amount of
telescope time granted, and the scope of the programme). Programme IDs
may be readily associated to published papers and to the abstract of the
proposals by entering it into the online form available at {\tt
  telbib.eso.org}. Table~\ref{Tab_IDs} is  organised as follows:
Column~1 gives the programme ID and col. 2 the last name of the
Prinicipal Investigator of the observing programme. Column~3 is a
brief description of the scope of the observing programme and nature
of the observed targets. Column~4 gives the amount of time allocated
to the proposal. Column~5 is the most frequently used grism in that
programme.

\subsubsection{Comments to individual programme IDs}\label{Sect_Comments}
Programmes 060.A-9203 and 060.A-9800 were not granted to users but
belonged to Paranal SCIOPS team for ordinary calibrations and for
technical tests, including some polarimetric tests on magnetic
stars. For instance, the first observation of a magnetic Ap star
(HD\,94660), used as a proof-of-concept for the method of
Sect~\ref{Sect_Bz_Meas} \citep{Bagetal02}, was made on 2001-03-23 under programme ID
060.A-9203 (although telescope time was officially granted as DDT).

This catalogue contains little or no reduced data for the programmes IDs
 65.H-0293 (PI=Jordan), 
 65.P-0701 (PI=Wagner), 
 66.D-0128 (PI=Appenzeller),
072.D-0736 (PI=O'Brien), 
072.D-0119 (PI=Marsh),
073.D-0322 and 082.D-0695 (PI=Reinsch) and 
277.D-5034 (PI=Greiner),
for various different reasons, e.g.\ because data were taken in
multi-object mode (073.D-0322, 082.D-0695) and could not be reduced by
the automatic pipeline, or because the S/N ratio was extremely low (e.g.\ 277.D-05034).
Most of data obtained under programme IDs
080.D-0521 (PI=Kawka) 082.D-0736 (PI=Vornanen) could not be used to
measure \bz\, through the least-square technique of
Sect.~\ref{Sect_Bz_Meas}, but intensity spectra are still made
available (see Sect.~\ref{Sect_Intensity}).

Programme IDs 68.D-0403, 70.D-0352, 073.D-0498, 272.D-5026 and
074.D-0488 (PI=Bagnulo) do contain many observations obtained in fims
mode, and were not properly reduced by our pipeline. However, it was
possible to re-use an older data reduction carried out with IRAF
routines by \citet{Bagetal06}. From this old data reduction we took
the extracted wavelength calibrated beams and we recombined them to
obtain the \pv\ and \nv\ spectra, and measured the \bz\ and \nz\ values
as explained in Sect.~\ref{Sect_Spectra} and \ref{Sect_Bz_Meas}.

\subsection{The entries of the catalogue}\label{Sect_Entries}
The general catalogue includes the following entries; the number in parentheses
correpond to the fields (columns) of the catalogue.\\
1-7) The coordinates RA and DEC of the frame fits-headers. These may not correspond
exactly to Simbad coordinates, but in the large majority of the cases they allow an unambiguous
identification of the target.\\
8) The star identifier.\\
9) The star classification as explained in Sect.~\ref{Sect_Classification}.\\
10) The programme ID.\\
11) The epoch of the observations (at the mid-time of the exposure series), expressed in Modified
Julian Date.\\
12-13) Same as 11), but expressed in calendar date and UT.\\
14) The total exposure time in seconds.\\
15) The number of frames used for the field determination.\\
16) The grism used.\\
17) The slit width in arcsec.\\
18) The spectral resolution measured on the arc lamp in a spectral line situated
approximately in the spectrum centre.\\
19--20) The wavelength spectral range (blue and red ends, in \AA).\\
21) The S/N ratio per \AA\ calculated as the median of the 100 highest pixels of
the spectrum (but excluding emission lines).\\
22) The centre of the wavelength interval where the S/N ratio peaks.\\
23--25) The mean longitudinal field measurement \bz\ from Balmer lines with its
error bar, and the corresponding reduced $\chi^2$; if
H Balmer lines are absent in the spectrum, the field value and its error bar
are set to zero. \\
26--28) Same as 23--25) for the field measured from the null profiles, \nz.
For the observing series including only one pair of exposures, null fields 
values and their error bars are set to zero.\\
29--34) Same as 23--28) for metal lines; if metal lines are absent in the spectrum,
all these columns contain zero values.\\
35--40) Same as 23--28) for both H Balmer and metal lines; if H Balmer lines are
absent in the spectrum, these columns contain the same values as cols.\ 29--34).
Conversely, if metal lines are absent from the spectrum, these columns contain
the same values as cols.\ 23--28).\\
41) The name of a downloadable gzipped tar file that contains the intensity spectra described in
Sect.~\ref{Sect_Intensity}. \\

There exist catalogue entries that do not include magnetic field
determinations at all, which correspond to cases where the field is
unmeasurable either because the spectrum has only (non-photospheric)
emission lines, or because the spectrum is featureless, or because it
is formed outside of the Zeeman regime, and Eq.~(\ref{EqBz}) does not
provide its estimate. These entries are still kept in the catalogue
because it is still useful to know that a certain observing series has
not been overlooked at. Furthermore the intensity spectrum is still
potentially useful and available in the archive of reduced data, and a
catalogue entry helps to identify the essential information about the
observations.

\subsection{Abridged printed version of the catalogue}
Table~\ref{Tab_Cat} is an abridged version of the catalogue of the
FORS magnetic field measurements and includes only the star name, the
stellar classification, the MJD of the mid of the observation, the
grism used, the \bz\ and \nz\ values in G with their error bars and
corresponding reduced $\chi^2$. When no magnetic field measurement is
available, a blank is left in the corresponding columns. 

The last column of this Table is a three-character flag that helps to
identify possible field detection. Each character reflects the results
of the analysis carried out on the H Balmer lines, on the metal lines,
and on the combination of the two sets of lines, respectively. An ``n''
means that the absolute field values was $< 3\,\sbz$, a ``d''
corresponds to the cases where $3 \le \vert\bz\vert/\sbz\ \le 5$,
and a ``D'' when the field was detected at more than 5\,$\sigma$
level.  The reason we assign a weaker detection certainty (d) to
detections at the 3 to 5\,$\sigma$ level than we do to detections at
higher significance (D) is because of the problem of “occasional
outlier” detections that has been clearly identified in FORS1 data
\citep{Bagetal12}. 

Our final assessment whether the star is really magnetic or not is
given by the ``M'' flag in the classification. Since this assessment
is often also based on measurements obtained with other instruments
and on other observing dates, it is quite possible that a star
classified as magnetic in col.~2 has a flag ``nnn'' in the last
column. Conversely, ``d'' flags (or even ``D'' flags) may be associated
with stars that have not received the classification of magnetic stars
in col.~2. This may happen for three reasons: either we think that the
detection might be real, but that it needs be supported by further data
(e.g.\ the case of the M giant star HD\,298045), or we believe that the
signal we measure is spurious, and/or the field detection has not been
confirmed by further observations obtained with FORS1 itself or other
instruments. Most of these spurious or dubious cases have been
discussed by \citet{Bagetal12}.

Finally, a dash (``-'') in the last column means that the
corresponding part of the spectrum was not used to measure the field
(in fact, there are cases where no field measurement was attempted, as
explained in Sect.\ref{Sect_Entries}).

\subsection{The archive of intensity spectra}\label{Sect_Intensity}
\begin{table*}
\caption{\label{Tab_Textfile} Example of the content of the input file
{\tt STARNAME\_PID\_III.X-JJJJ\_MJD\_nnnnn.mmm}: here we consider
{\tt HD190073\_PID\_081.C-0410\_MJD\_54609.411} (the file structure is explained
in the text). From the file list one
can infer that last two exposures of the observing series have been
dropped (either because the observing series was interrupted, or 
because the frames were discarded due to some problem, e.g.\ saturation).
}
\begin{center}
\begin{tabular}{lrrrr}
FORS1.2008-05-23T09:36:34.552.fits &  180 & 315.0 &   1/  8 & 200176637\\
FORS1.2008-05-23T09:41:07.012.fits &  200 &  45.0 &   2/  8 & 200176637\\
FORS1.2008-05-23T09:45:27.830.fits &  250 &  45.0 &   3/  8 & 200176637\\
FORS1.2008-05-23T09:51:10.408.fits &  250 & 315.0 &   4/  8 & 200176637\\
FORS1.2008-05-23T09:56:21.302.fits &  250 & 315.0 &   5/  8 & 200176637\\
FORS1.2008-05-23T10:02:03.889.fits &  250 &  45.0 &   6/  8 & 200176637\\
\end{tabular}\\
\hspace*{-7.25cm}GRIS\_600B 0.40 1797\\
\hspace*{-4.2cm}Norma III 4136x4096 1x1 200Kps/low\_gain
\end{center}
\end{table*}
In the course of measuring the magnetic field strength from the
circularly polarised spectra of FORS1, we have obtained the
(uncalibrated) intensity ($I$) spectra. Since FORS is a single order
spectrograph, these $I$ spectra, even without flux calibration,
provide potentially useful profiles of broad spectral lines such as
those of H and He that are difficult to recover with accuracy from
high-dispersion spectra derived from cross-dispersed instruments. The
$I$ spectra also illustrate clearly the overall shape of the detected
spectral flux (convolved with the instrument+telescope transmission function)
of each star observed. Because these $I$ spectra could
potentially be useful for a wide range of projects, we have made them
available at CDS, and, temporarily, at {\tt http://star.arm.ac.uk/FORS/}

In the ESO archive, each frame is identified by a name that
refers to the instrument and the instant when an exposure was
started. In case of FORS1 data:\\
{\tt FORS1.YYYY-MM-DDThh-mm-ss.xxx.fits}\\ 
where {\tt YYYY-MM-DD} refers to the year, month and day of the
observation, and {\tt hh-mm-ss.xxx} the hour, minute and second (with
millisecond precision) when shutter was open for the
observation (UT). We note that files produced until the end of period 67 
were called {\tt FORS.YYYY-MM-DDThh-mm-ss.xxx.fits}. 

In our context it is useful to group all together the frames of the
observing series that have been used to obtain a certain magnetic
field measurement.  Therefore, for each entry of our catalogue, we
have produced a tarball named {\tt
  STARNAME\_PID\_III.X-JJJJ\_MJD\_nnnnn.mmm}.tar where {\tt STARNAME}
is the star name, {\tt III.X-JJJJ} is the programme ID and {\tt
  nnnnn.mmm} is the Modified Julian Date of the observation. Very few
objects oberved in multi-object mode could not be identified
\citep[see][]{Bagetal06}. In these cases, for the instead of {\tt
  STARNAME} we used {\tt RAhh\_mm\_ss.s} where hh:mm:ss.s is the RA of
the centre of the slit read in the fits-headers.  This way, each
tarball is unambiguously associated with each entry of the catalogue and
Table~\ref{Tab_Cat}.

Each tarball includes an ASCII file with the same name as the tarball
itself (without the extension .tar) which contains the list of
original frames used for the field determination. In this file, each
filename is followed by: the exposure time in seconds of each
individual frame (fits-header keyword {\tt EXPTIME}); the position
angle of the retarder waveplate with respect to the parallel beam of
the Wollaston prism (fits-header keyword {\tt
  INS.RETA4.ROT})\footnote{Older FORS1 data did not include this
  keyword, in which case it was calculated as the difference between
  the fits-header keywords {\tt ADA.POSANG} and {\tt INS.RETA4.POSANG}
  which give the position angles of the instrument and of the retarder
  waveplate, respectively, with respect to the north celestial
  meridian.}; the exposure number, and the total number of exposures
in each OB template, and the OB number, given by fits-keywords {\tt
  NEXP} {\tt EXP} and {\tt OB}, respectively. The file name list is
followed by two lines with the grism name, the slit width, the
spectral resolution, the detector name and the readout mode. An
example of such file is given in Table~\ref{Tab_Textfile}.

The tarball includes an ASCII file for each frame of the observing
series used to determine \bz. This file is identified by the same name
as the original frame archive name, having replaced the extension {\tt
  .fits} with {\tt .prof}. In case of observations obtained in
multi-object mode, the name of the .prof file refers also to the
slitlet number where the star was centred, e.g., {\tt
  FORS1.YYYY-MM-DDThh-mm-ss.xxx\_Syy.fits}, where yy may be 02, 04,
\ldots, 18. Each of these .prof files have three columns: wavelength
in \AA\ (col. 1), flux and flux error in ADUs (cols.\ 2 and 3,
respectively).

\section{Comparison with previously published field values}\label{Sect_Comp}
It is of interest to compare the \bz\ values obtained from the current
suite of reduction programs with results published in the literature
and obtained from the same datasets.

\subsection{Data on magnetic Ap/Bp stars}
A first comparison may be made with the Ap star field strength values
obtained for open cluster Ap/Bp candidates that are described by
\citet{Bagetal06}. These measurements were made using a combination
of Balmer and metallic line in order to maximise sensitivity to weak
fields in these often faint stars. Field measurements made for this
observing programme are found to have uncertainties that are rather similar to
those in the present catalogue. For stars in which no field was
detected, in general the result is still a null detection, although
the actual value reported has often changed by an amount of the order
of 1\,$\sigma$ or even more, while still remaining a null
detection. For stars with easily detectable fields, the values of the
uncertainties are not greatly different from the present data, but the
actual \bz\ values reported may differ from the current ones by 10 or
even 20\,\%.

Another comparison may be made with the data used by \citet{KocBag06}
to study the evolution of magnetic field strength with age among
bright field Ap/Bp stars. These data differ from the open cluster field
strengths in that the stars observed are generally a few magnitudes
brighter, and so the S/N ratio of the measurements are
higher. Field strengths were measured using only the
Balmer lines, as this provided adequate precision for their
project. Because the printed version of the catalogue presented here
has field values derived from both metal and H lines, the catalogue
uncertainties may be as much as two times smaller than those of
\citet{KocBag06}, and range up to similar or slightly
larger values in stars in which the metallic spectrum contributed
little useful information. Again, the actual values of \bz\ in the
present catalogue for stars in which the field is easily detected may
differ from the earlier values by 10 or 20\,\%, in this case at least
mainly because the field is not determined using the same lines in the
two datasets.

\subsection{White dwarfs}
We have also compared the catalogue to the field measurements of DA
white dwarfs described by \citet{Lanetal12b}. Since this publication
was based on a method very similar to that adopted for this catalogue, the field values
and uncertainties are generally very similar in the two places. The
main exception concerns the two (null) field measurement of the white
dwarf WD\,1334$-$678 that were included (by mistake) in their online
Table~3 of all FORS measurements of potential kG field DA white
dwarfs. We discovered that the actual target of that measurement is in
fact an anonymous G star rather than the white dwarf, and that the
white dwarf is actually more than 1 arcmin away from the fits-header
coordinates. In this catalogue, the observations have been assigned to
the correct star (identified by the fits-header coordinates), and no field value is included in the
catalogue, although the $I$ spectrum is being made available.

For all the remaning stars, small field value differences are present,
and due to the slightly different version of the algorithm that we
have adopted for this catalogue.

\subsection{Other stars}
A thorough comparison of the results in the catalogue with reported
magnetic field discoveries in non-Ap/Bp stars made by other groups
from FORS1 observations has been carried out by \citet{Bagetal12}. The
general result of this comparison was that many of the reported
discoveries are erroneous or at least unsupported by a revision of the
original FORS1 data.

In Sect.~5 of \citet{Bagetal12} we highligted a number of cases where
our pipeline-based data reduction was unsatisfactory. We have
re-addressed these cases, sometimes using a ``hand-made'' data
reduction, and found the conclusions described in the remaining part
of this Sect., which addresses a dozen very specific cases.

Table~5 of \citet{Bagetal12} reported no detection but very large
error bars for the observations of Be star HD\,148184 on MJD=53532.224
and 53862.380. Our new reduction produces much smaller error bars, and
still does not confirm the detection previously reported in the
literature.

Section 5.2 of \citet{Bagetal12} reported apparently significant but
very suspicious field detections for four classical Be stars: an
observation of HD\,181615=HD\,181616, one observation of HD\,56014,
two observations of HD\,209409 (in which the original observers did
not report any significant fields), and one observation of
HD~224686. Our new data reduction has fixed all these problems and no
detection is reported, fully confirming the claim by \citet{Bagetal12}
that from FORS data there is no evidence for magnetic fields in any
classical Be star.

Similar problems affected three slowly pulsating B (SPB) stars
\citep[see comments in Sect.~5.4 of][]{Bagetal12}.

For the \bz\ measurement of the SPB star HD\,53921 obtained on
MJD=52999.137, \citet{Bagetal12} reported a field detection with the
opposite sign to that measured by \citet{Hubetal06}, and for the
measurements obtained at MJD=53630.401 and 53631.408,
\citet{Bagetal12} reported error bars five times higher than
previously reported by \citet{Hubetal06}, with no significant
detection in these two datasets (however, Bagnulo et al.\ 2012
confirmed that the star is magnetic based on HARPSpol
measurements). With our new reduction we are able to confirm the field
detection on MJD=52999.137 with a positive sign, and we have gotten
rid of additional noise in the remaning two measurements, confirming
the 5\,$\sigma$ detections reported by \citet{Hubetal06}.

\citet{Bagetal12} reported an unsatisfactory reduction for the SPB
star HD\,152511 on MJD=54609.433 due to seeing conditions. Our
revision of this dataset lead to much better results, which are
consistent with those of \citet{Hubetal09}. Therefore we confirm all
three detections reported by \citet{Hubetal09} instead of only two as
reported by \citet{Bagetal12}.

For the observations of the SPB star HD\,28114 obtained on MJD=54106.091
\citet{Bagetal12} obtained a larger noise than was previously
published by \citet{Hubetal09}, and a field detection based only on a
signal that appears in the highest-order H Balmer lines. Our new
reduction has a higher S/N ratio, but we still get the same
suspicious signal only on the highest-order H Balmer lines, and no
credible field detection.

Section~5.4 of \citet{Bagetal12} conclude that six reported detections
of a field in the $\beta$\,Cep star HD\,16582 had decreased below the
3\,$\sigma$ significance limit, although one measurement not
originally claimed as detection, on MJD=54343.259, has risen to become
an apparently significant detection. This detection has disappeared in
our new reduction, highlighting once again how the reliability of
marginal discoveries may crucially depend on data-reduction.

\section{FORS detections of stellar magnetic fields}
With the complete dataset of magnetic measurements obtained with
FORS1, we can take stock of the achievements of this instrument, and
assess its strengths and weaknesses.

We consider first FORS1 measurements of Ap/Bp stars. A large number of
such stars have been observed, both in open clusters and in the field,
and in the context of wide-ranging surveys as well as for studies of
single objects. For such stars, the detection rate is reasonably high:
if the star is securely identified as an Ap/Bp star (by specific
chemical peculiariies, or by appropriate values of the photometric
Maitzen $\Delta a$ or Geneva $Z$ peculiarity parameters), then the
likelihood of clear detection of a longitudinal field is around
50\,\%. This result occurs because in practice the main surveys have
been able to achieve measurement uncertaintites of the order of
50–-100\,G, while the fields to be detected are typically several
hundred G. Thus a measurement with $\bz/\sbz$ ratio of order ten is often
obtained, a value large enough to clearly establish the presence and
amplitude of a field in spite of the excess noise that can sometimes
trouble FORS1 measurements. It is clear that FORS is extremely
powerful as a tool to search for fields in such stars, down to
magnitudes fainter than $V \sim 10$, and is perfectly capable of
detecting kG fields in stars as faint as $V=13$ or 14.

The FORS1 Ap stars data have recently been discussed by
\citet{Lanetal14}, who studied the general usefulness of FORS for
systematic studies of individual Ap stars, and concluded that (apart
from occasional outliers) the instrument furnishes data of high
quality and consistency. However, only a few of the observing
programmes carried out on FORS1 have focused on this kind of problem.

With the discovery that WD\,446-789, WD\,2105-820, WD\,2359-434 (and
perhaps also WD\,1105-048) host a magnetic field, the FORS1 surveys of
white dwarfs have opened a new stream of research, i.e.\ systematic
investigations of weak field (10\,kG or less) in degenerate stars
\citep{Aznetal04,Lanetal12b}, which definitely justify the use of a
telescope with an 8\,m size mirror. FORS1 has played also an important
role in the study of faint but stronger magnetic 
white dwarfs \citep{KawVen12} and was used for the discovery 
of circular polarisation in the continuum and in the molecular
bands of a DQ white dwarf \citep{Voretal10}.

When we look at the large number of observations of stars other than
magnetic Ap/Bp stars, and the very small number of secure field
detections, it is clear that most of the stellar magnetism programmes
carried out on FORS1 of non-Ap/Bp stars were searches for fields in
individual objects, or surveys of various classes of stars for
detectable fields. The projects carried out have included large
surveys of such star classes as Herbig Ae/Be stars, O stars, slowly
pulsating B stars, $\beta$~Cep B star pulsators, classical Be stars,
normal B stars, and white dwarfs. A number of individual objects have
also been studied.

However, from the $\sim 1000$ measurements carried out on stars other
than Ap stars and white dwarfs, we have only a few clear field
detections, namely 
the pre-main-sequence star HD\,101412, the $\beta$~Cep variable
HD~46328, the SPB stars HD\,53921 and HD\,152511, and the rapidly
rotating star FK~Com. In addition, a number of 3--6\,$\sigma$ FORS1
detections have been reported in the literature (some of which still
present in this catalogue). As thoroughly discussed by
\citep{Bagetal12}, some of these detections have been proved by
subsequent monitoring with FORS2 and ESPaDOnS to be real (e.g.\ the
Of?p star HD\,148937), but many of them have not been confirmed by our
reduction, or by observations with other instruments, and are probably
spurious. A small number of cases would deserve further investigations
(e.g.\ the SPB star HD\,138769, and the M giant star HD\,298045).

The null results obtained in the various surveys of stars other than
Ap stars are certainly valuable for setting upper limits on possible
fields, e.g.\ on RR Lyrae stars \citep{KolBag09}, on hot subdwarfs
\citep[see][and references therein]{Lanetal12a}, and on the central stars
of planetary nebulae \citep{Leoetal11,Joretal12}.
These results are often useful for constraining possible
theoretical models of various kinds.  The extremely low detection rate
is mainly a consequence of the rarity in hot stars of fields that are
large enough to be clearly detectable with FORS1 spectropolarimetry;
such stars have a frequency of occurrence of 10\,\% or less. It may
also be a consequence of a tendency of both proposers and the OPC to
prefer the 8\,m telescope for exploration of new fields rather
than systematic study of individual fields found. In any case, the low
yield suggests that surveys with FORS need to be Large Projects, and
that otherwise the instrument should focus more on systematic study of
single objects in which the field is known to be large enough to be
studied at useful S/N ratio with FORS.

\section{Conclusions}
This paper is the concluding work in a series that
started with the first demonstration that FORS1 could be used
effectively for magnetic field measurements of main sequence stars
\citep{Bagetal02}. This was followed by several years
during which FORS1 (and, later, the twin instrument FORS2)
was widely used for observations of many classes of
stars. After some years it became apparent that a general discussion
of the analysis of data obtained with dual-beam spectropolarimeters
similar to that of FORS1 would be of considerable value to the
community. This led to a paper presenting the fundamental ideas of
the beam-swapping technique, and its application to night time
astronomy \citep{Bagetal09}.

Further use of FORS1 led to announcements of numerous field
discoveries at the 3 to 5\,$\sigma$ level. When these results were not
confirmed by our own reduction of the data and/or were contradicted by
observations with other spectropolarimeters such as ESPaDOnS, it
became apparent (1) that correct treatment of FORS1 data is less
obvious than it appears at first sight, especially in the regime of
small uncertainties and marginal field detections, and (2) that the
instrument itself may be subject to small drifts and flexures that
lead to erroneous data in a small fraction of cases. These issues were
discussed in considerable detail by \citet{Bagetal12}, who developed
a suite of programs allowing the entire dataset of all magnetic
observations obtained with FORS1 to be reduced together, with a
variety of options. This tool made it possible to show that modest and
reasonable improvements to the data reduction process (for example,
optimal choice of a clipping algorithm to deal with cosmic ray events)
could change field detections into non-detections (and occasionally
vice versa).

\citet{Bagetal12} also clearly identifed the occurrence of occasional
outliers in FORS1 data, in which a single observation inconsistent
with others obtained with FORS1/2, could occur. An important source of
such occasional outliers, and more generally of excess noise in
FORS1/2 field measurements, was identified by \citet{Bagetal13}, who
showed that small shifts in line position may occur in FORS data,
probably as a result of small flexures (including
spectral shifts following the rotation of the retarder waveplate) and of
seeing, and that these shifts can lead to spurious field
detections.

The present paper builds on the accumulated experience and
experiments represented by these earlier works,
also including the detailed
discussion of the large body of field measurements of AP/Bp stars by
\citet{Lanetal14}.
 It does not, of course
present {\it the} definitive reduction of the FORS1 magnetic dataset;
as discussed in the earlier papers, several different choices made
during data reduction are equally reasonable, but lead to somewhat
different results. However, the reductions leading to the present
catalogue are based on reasonable choices, applied in a
consistent way to all the FORS1 field measurements, and they do rest
on an understanding of, and correction of, small errors or poor 
choices made to treat data in the past.

The results compiled in this catalogue show that FORS is a very
powerful instrument for (longitudinal) field measurement of both
non-degenerate and degenerate stars. In the very best cases (for stars
with rich spectra of deep lines), uncertainties of as low as 20\,G or
so can be achieved. Because of
the outlier problem, new field detections should be repeated multiple
times and/or made at the significance level of at least 5 or
$6\sigma$. More typically, FORS can achieve realistic uncertainties of
the order of 50--100~G for a wide range of stars down to $V$ magnitude
of 10 to 12, and $\sim 500$\,G uncertainties for white dwarfs of
magnitude 12 or 13.  The instrument is particularly well-suited to
measurement of fields that can be detected at several $\sigma$
level.

It should be borne in mind, however, that FORS makes very inefficient
use of telescope time in observations of stars brighter than $V \sim
7$, for which the readout time of the CCD is larger than the
integration time of each subexposure. This is the origin of the
striking difference between shutter time and allocated time mentioned
at the beginning of Sect.~6. There may be times when FORS may still be
the best instrument for the job, but it is used most efficiently on
stars for which subexposure times of some or many minutes are needed.

The relevance of most of the considerations made in this series of
papers are not limited to the FORS1 instrument, but should be taken
into account when using/designing other spectro-polarimeters.

\begin{acknowledgements}
This paper is based on observations made with ESO telescopes at the La Silla
Paranal Observatory under the programme IDs listed in Table~\ref{Tab_IDs}, and
made available through the ESO archive. LF acknowledges
support from the Alexander von Humboldt Foundation.  Work on this
project by JDL has been supported by the Natural Sciences and
Engineering Research Council of Canada. We thank the referee
P.~Petit, for a careful review of the manuscript.
\end{acknowledgements}

\newpage
\begin{small}
\begin{longtable}{llrlrr@{\,$\pm$\,}lrr@{\,$\pm$\,}lrc}
\caption{\label{Tab_Cat} Magnetic field and null
field are calculated from the combination of H Balmer and metal lines.} \\
\hline\hline
Star                                 &   
Classification                       &   
\multicolumn{1}{c}{Prog. ID}         &   
\multicolumn{1}{c}{MJD}              &   
grism                                &   
\multicolumn{2}{c}{\bz\ (G)}         &   
\multicolumn{1}{r}{$\chi^2/\nu$}     &   
\multicolumn{2}{c}{\nz\ (G)}         &   
\multicolumn{1}{r}{$\chi^2/\nu$}     &   
\multicolumn{1}{r}{HmT}            \\   
\hline
\endfirsthead
\caption{continued.} \\
Star                                 &  
Classification                       &  
\multicolumn{1}{c}{Prog. ID}         &  
\multicolumn{1}{c}{MJD}              &  
grism                                &  
\multicolumn{2}{c}{\bz\ (G)}         &  
\multicolumn{1}{r}{$\chi^2/\nu$}     &  
\multicolumn{2}{c}{\nz\ (G)}         &  
\multicolumn{1}{r}{$\chi^2/\nu$}     &  
\multicolumn{1}{r}{HmT}            \\   
\hline
\endhead
\hline
\endfoot
WD\,2359-434           &WD:DA5:M       &070.D-0259&52583.025&600B &$  3692$& 842&  1.02&$   496$& 853&  1.07&d-d\\
WD\,2359-434           &WD:DA5:M       &070.D-0259&52608.056&600B &$  3144$& 528&  1.02&$   663$& 483&  0.86&D-D\\
HD\,225041             &MS:A:RR        &082.D-0342&54781.102&1200B&$     4$&  32&  1.49&$    12$&  25&  0.96&nnn\\
HD\,225264             &MS:A1:AP       &073.D-0498&53217.409&600B &$   -26$&  34&  0.90&$   -22$&  31&  0.79&nnn\\
HD\,358                &MS:B9:HGMN     &071.D-0308&52910.092&600B &$  -304$& 123&  1.08&$    31$& 120&  1.02&nnn\\
HD\,358                &MS:B9:HGMN     &072.D-0377&52963.020&600B &$   139$& 101&  1.00&$  -150$& 102&  1.01&nnn\\
HD\,358                &MS:B9:HGMN     &075.D-0295&53519.448&1200g&$   -88$&  97&  2.23&$   194$& 107&  2.23&nnn\\
HD\,358                &MS:B9:HGMN     &075.D-0295&53629.286&1200g&$    23$&  30&  1.21&$    23$&  29&  0.95&nnn\\
\hline
\end{longtable}
\end{small}

\end{document}